\documentclass[12pt]{article}
\usepackage{multirow}
\usepackage{array}

\usepackage[T1]{fontenc}
\usepackage{geometry}
\usepackage{threeparttable}
\usepackage{graphicx}
\usepackage{epstopdf}

\usepackage{color}
\usepackage{longtable}
\usepackage{amsmath}
\usepackage{amsthm}
\usepackage{amssymb}
\usepackage{bm}
\usepackage{comment}
\usepackage{setspace}
\usepackage{esint}
\usepackage{subcaption}
\usepackage{enumerate}
\usepackage[authoryear]{natbib}
\usepackage{paralist}
\usepackage[unicode=true,bookmarks=true,bookmarksnumbered=false,bookmarksopen=false,
breaklinks=true,pdfborder={0 0 1},backref=section,colorlinks=true]
{hyperref}
\hypersetup{linkcolor=blue,urlcolor=blue,citecolor=blue,setpagesize=false,linktocpage}
\renewcommand{\theequation}{\arabic{equation}}
	
\makeatletter
\theoremstyle{plain}

\theoremstyle{plain}

\usepackage[margin=28pt,font=small,labelsep=endash]{caption}
\usepackage{amsfonts}
\usepackage{authblk}
\providecommand{\U}[1]{\protect\rule{.1in}{.1in}}
\makeatother
\providecommand{\assumptionname}{Assumption}
\providecommand{\propositionname}{Proposition}
\makeatother
\providecommand{\assumptionname}{Assumption}
\providecommand{\lemmaname}{Lemma}

\begin{document}
	\title{A Deep Learning Framework for Medium-Term Covariance Forecasting in Multi-Asset Portfolios\thanks{This work was carried out within the scope of the research project funded by Fundação para a Ciência e a Tecnologia, grant number 2023.01070.BD.}}
	
	\author{Pedro Reis\thanks{University of Porto, School of Economics and Management and Center for Economics and Finance (cef.up), R. Dr Roberto Frias 46, 4200-464 Porto, Portugal; and INESCTEC, Campus da Faculdade de Engenharia da Universidade do Porto, R. Dr. Roberto Frias, 4200-465 Porto, Portugal. Email: pdreis1988@gmail.com},
		\ Ana Paula Serra\thanks{University of Porto, School of Economics and Management and Center for Economics and Finance (cef.up), R. Dr Roberto Frias 46, 4200-464 Porto, Portugal.} \ and João Gama\thanks{INESCTEC, Campus da Faculdade de Engenharia da Universidade do Porto, R. Dr. Roberto Frias, 4200-465 Porto, Portugal.}}
	
	\date{\vspace{-5ex}}
	\maketitle
	\clearpage
	\begin{abstract}
		Accurate covariance forecasting is central to portfolio allocation, risk management, and asset pricing, yet many existing methods struggle at medium-term horizons, where shifting market regimes and slower dynamics predominate. We propose a deep learning framework that combines three-dimensional convolutional neural networks, bidirectional long short-term memory layers, and multi-head attention to capture complex spatio-temporal dependencies. Using daily data on 14 exchange-traded funds from 2017 through 2023, we find that our model reduces Euclidean and Frobenius distance metrics by up to 20\% relative to classical benchmarks (e.g., shrinkage and GARCH approaches) and remains robust across distinct market regimes. Our portfolio experiments demonstrate significant economic value through lower volatility and moderate turnover. These findings highlight the potential of advanced deep learning architectures to improve medium-term covariance forecasts, offering practical benefits for institutional investors and risk managers. 
	\end{abstract}
	
	\clearpage
	\section{Introduction} 
	\label{sec:intro}
	
	Covariance matrix forecast is central in modern finance. It influences portfolio construction, risk management, and asset pricing models. \citet{Markowitz1952} introduced mean-variance optimisation, highlighting how correlation among assets shapes the risk-return trade-off. Beyond portfolio choice, covariance forecast underpins factor-based models, hedging strategies, and even debates on market efficiency, where accurate correlation forecasts can drive risk control and potential outperformance.
	
	Early studies found that even simple correlation structures could improve capital allocation. For instance, \citet{Elton1973} suggested using constant-correlation estimates over historical windows for a decade. Their work showed that stable or slowly changing correlation patterns can help avoid distortions caused by transient market fluctuations.
	
	Over time, multiple research strands have focused on short-horizon covariance forecasts. High-frequency datasets, such as intraday prices, enable detailed volatility analyses, often using generalised autoregressive conditional heteroskedasticity (GARCH) based models. Constant \citep{Bollerslev1990} and dynamic conditional correlation \citep{Engle2002} specifications remain common benchmarks, although \citet{Symitsi2018} argue that little fundamental progress has been made recently. Extensions include applying the heterogeneous autoregressive model \citep{Corsi2009} in a multivariate setting through Cholesky decomposition\footnote{Cholesky decomposition is a method that breaks down a positive definite matrix into the product of a lower triangular matrix and its transpose.} \citep{chiriac2011}. However, the Cholesky decomposition models outperform DRD-based\footnote{Dynamic Relationship/Dependence (DRD) models refer to time-varying models that separate covariance into volatility and correlation components forecasts.} \citep{Bollerslevetal2018}.
	
	However, GARCH-type approaches struggle with large cross-sections, falling prey to the curse of dimensionality \citep{Engleetal2019}. Tools like principal component analysis and random matrix theory \citep{Ledoit2003, Laloux2000} have been effective for large datasets. Recent advances, such as dynamic conditional correlation with nonlinear shrinkage \citep{Engleetal2019}, perform well overall, as per \citet{Nardetal2024}. Still, most of these methods focus on very short-term horizons, often just one step ahead or one day ahead, and rely on high-frequency data.
	
	Long-horizon forecasts matter for institutional investors, such as pension and investment funds. They rebalance portfolios over weeks or months. Therefore, a one-step forecast is insufficient for practical decision-making \citep{DeNard2022}. Furthermore, slow-moving trends, such as equity–bond correlation shifts, can be overlooked if one relies solely on high-frequency econometric models. \citet{Sandoval2012} highlights how regime shifts or significant structural changes often go undetected in classical specifications, raising the risk of large drawdowns and poor capital allocation.
	
	For longer-term horizons, naive and factor models are more common \citep{Ledoit2004, DeNard2022}. Factor models can provide insightful forecasts but may not generalise well across diverse portfolios, especially when different portfolios are affected by many different factors. These methods are typically stable but can underperform when major shifts occur. 
	
	Naive approaches assume that recent covariance estimates strongly predict future covariances as they follow a Markov process. However, these tend to be noisy when the number of assets tends to be close to the number of observations ($N \approx T$). For this purpose, shrinkage techniques reduce noise by blending the sample covariance matrix with structured targets, particularly when the cross-section of assets is large \citep{Ledoit2022}.
	
	While these methods rely on advanced statistical and econometric methods, we have seen a recent surge in machine learning methods that can also handle non-linearities and high dimensionality with more accurate forecast. \citet{Gu2020} and \citet{Kim2023} use techniques like random forests, support vector machines, and generative adversarial networks, while \citet{Zhang2024} applies graph neural networks to capture volatility spillovers for covariance matrix forecast. However, most of these solutions still emphasise short-term horizons.
	
	Attention mechanisms \citep{Vaswani2017} offer a promising direction by weighting key observations more heavily. Some studies apply these techniques in finance but still focus on short-horizon tasks \citep{Nazareth2023, Olorunnimbe2023}. In principle, attention-based models could capture the slow-moving, cyclical factors and abrupt shifts that define medium- and long-horizon covariance forecasts, realising this potential. However, it requires integrating sophisticated data-driven architectures with the economic logic of excess returns.
	
	Recent efforts to extend these approaches to multi-asset portfolios have been limited. While many studies handle large cross-sections of stocks or focus on single-asset classes \citep{Reis2024}, fewer examine the combined dynamics across multiple asset classes in an integrated framework. However, institutional investors routinely diversify across equities, bonds, alternatives, and derivatives, underscoring the need for robust, multi-asset covariance forecasting models. In particular, capturing time-varying cross-asset correlations over medium-term horizons is a non-trivial challenge that remains under-explored in much of the existing literature.
	
	Two key research gaps remain. First, most existing studies focus on short- or long-term horizons, leaving a relative void in medium-term covariance forecasting. This gap is particularly salient for multi-asset portfolios, where cross-asset correlations over multi-week or multi-month windows can significantly influence asset allocation and risk management \citep{DeNard2021}. 
	Second, there is limited empirical evidence on whether using \emph{raw} returns versus \emph{excess} returns leads to systematically different covariance forecasts, especially over these medium-term horizons. Economic theory suggests that underlying drivers of asset co-movements could be more accurately captured through \emph{excess} returns \citep{Fama1993}, i.e. by removing the risk‑free component from returns. However, most machine learning and econometric approaches mostly default to \emph{raw} returns.
	
	Our study addresses these gaps by focusing on medium-term covariance forecasts in a multi-asset setting. Specifically, we propose a novel deep learning (DL) model that captures spatio-temporal correlations while remaining robust to structural changes. We benchmark against classical methods and demonstrate consistent gains in forecast accuracy under diverse market environments from 2017 to 2023. Our main contributions are as follows:
	\begin{itemize}
		\item We present an integrated 3D-convolution neural network (CNN) and bidirectional long-short-term memory (LSTM) architecture with multi-head attention that improves medium-term forecasting accuracy.
		\item We compare different return-processing specifications (including the treatment or omission of risk-free rates) and show that the proposed approach preserves its predictive edge.
		\item We evaluate the economic value of our forecasts in a global minimum-variance (GMV) portfolio, demonstrating significant variance reduction and stable turnover.
	\end{itemize}
	
	The rest of the paper is structured as follows. Section~\ref{sec:methodology} introduces the benchmark methods and our DL model. Section~\ref{sec:empirical} presents the main empirical findings. Section~\ref{sec:robust} reports robustness checks. Section~\ref{sec:portfolio} explores implications for portfolio management. Finally, Section~\ref{sec:conclusion} concludes.
	
	\section{Methodology}
	\label{sec:methodology}
	Let \(\mathbf{r}_{t}\) denote the \(N\)-dimensional row vector of excess daily returns for \(N\) assets at time \(t\).	Assuming time-varying mean excess returns, \(\mathbf{r}_{t}\) can be expressed as follows:
	\begin{equation}
		\mathbf{r}_{t} = \bm{\mu}_{t} + \mathbf{e}_{t} \label{2.2}
	\end{equation}
	where $\bm{\mu}_{t}$ is the \(N\)-dimensional vector of rolling means of the excess daily returns, and \(\mathbf{e}_{t}\) is the i.i.d. error terms with a normal distribution within each rolling window:
	\begin{equation}
		\mathbf{e}_{t} \sim N\left( \mathbf{0},\bm{\Sigma}_{t} \right) \label{2.3}
	\end{equation}
	
	Although empirical evidence suggests financial returns often exhibit heavy tails and non-zero skewness \citep{Cont2001}, we adopt the normality assumption because it is standard in many benchmark models (e.g., GARCH-based correlation models and Ledoit–Wolf shrinkage). However, we acknowledge that alternative distributions (e.g., Student-$t$ innovations, mixture models, or those explicitly modelling tail dependence) may better capture extreme events and skewness in practice\footnote{Extending our DL framework to accommodate such heavy-tailed or asymmetric distributions is an area for future research. Nevertheless, our empirical results in subsequent sections suggest that even a normality-based design can significantly enhance covariance forecasts, given the robustness of the data-driven layers in handling complex, nonlinear dependencies.} \citep{Hansen1994, Harvey1997}. 
	
	The rolling mean $\bm{\mu}_{t}$ and rolling standard deviation $\bm{\sigma}_{t}$ at time $t$ are calculated using a window of size $F$ as follows:
	
	\begin{equation}
		\bm{\mu}_{t} = \frac{1}{F}\sum_{k = t - F + 1}^{t}\mathbf{r}_{k}\label{2.4}
	\end{equation}
	\begin{equation}
		\bm{\sigma}_{t}^{2} = \frac{1}{F - 1}\sum_{k = t - F + 1}^{t} \left( \mathbf{r}_{k} - \bm{\mu}_{t} \right) \odot \left( \mathbf{r}_{k} - \bm{\mu}_{t} \right) \label{eq:2.5}
	\end{equation}
	where \(\mathbf{\odot}\) denotes the element-wise product. Therefore, the realised covariance matrix (\(\bm{\Sigma}_{t}\)) is an \(N \times N\) matrix with each element calculated as:
	\begin{equation}
		Cov_{t}\left( \mathbf{r}_{i},\mathbf{r}_{j} \right) = \frac{1}{F - 1}\sum_{k = t - F + 1}^{t}\left( r_{i,k} - \mu_{i,\, t} \right)\left( r_{j,k} - \mu_{j,\, t} \right) \label{2.6}
	\end{equation}
	
	Our goal is to estimate the conditional covariance matrix of the asset returns for the next \(F\) days based on the information available at time \(t\):
	\begin{equation}
		{\widehat{\mathbf{\Sigma}}}_{t + 1:t + F} = Cov\left( \mathbf{r}_{t + 1}, \mathbf{r}_{t + 2}, \ldots, \mathbf{r}_{t + F} \middle| \mathcal{F}_{t} \right) \label{2.7}
	\end{equation}
	where \(\mathcal{F}_{t}\) is the sigma-algebra representing all the information available at time \(t\).
	
	To estimate ${\widehat{\mathbf{\Sigma}}}_{t + 1:t + F}$, we compare 10 benchmark models divided into three groups (naive, de-noising and GARCH models) alongside our proposed model. A description of each model is shown below.
	
	\subsection{Naïve Models}
	\subsubsection{Naïve ($NA$)}
	A naive covariance forecasting approach is based on the lagged realised covariance. This model assumes that covariance is a Markov process, so the covariance matrix of the previous period is highly informative about the future covariance matrix. Under this model:
	
	\begin{equation}
		{\widehat{\mathbf{\Sigma}}}_{t + 1:t + F} = \mathbf{\Sigma}_{t - F:t} \label{2.8}
	\end{equation}
	
	This forecasting method is as simple as possible as it requires no optimisation.
	\subsubsection{Naïve full sample ($NA^F$)}
	Another naïve technique is to use the full sample instead of the rolling covariance matrix as our estimator.
	\begin{equation}
		{\widehat{\mathbf{\Sigma}}}_{t + 1:t + F} = \mathbf{\Sigma} \label{2.9}
	\end{equation}
	where $\mathbf{\Sigma}$ is the covariance matrix of our sample up until \(t\).
	
	\subsubsection{Exponential Weighted Moving Average ($EWMA$)}
	The EWMA employs exponentially decaying weights for the covariance matrix. The covariance matrix in the EWMA model is recursively computed as follows:
	\begin{equation}
		{\widehat{\mathbf{\Sigma}}}_{t + 1:t + F} = (1 - \eta)\mathbf{e}_{t}\mathbf{e}_{t}^{'} + \eta\mathbf{\Sigma}_{t - F:t} \label{2.10}
	\end{equation}
	where \(\mathbf{e}_{t}\) is the N-dimensional vector of the error terms of all assets from Eq.~\eqref{2.2}, at time \(t\), and \(\eta\) acts as a decay factor, confined within the range [0,1], dictating the rate at which the weights on past observations decrease. This factor has been approximated to be around 0.94 \citep{JPmorgan}.
	
	\subsection{De-noising Models}
	\subsubsection{Principal Component Analysis ($PCA$)}
	As noted by \citet{Ledoit2003}, empirical results show the usefulness of PCA in covariance matrix forecast. The PCA is computed as follows:
	\begin{equation}
		{\widehat{\mathbf{\Sigma}}}_{t + 1:t + F} = \mathbf{U\Lambda}\mathbf{U}^{\top} \label{2.20}
	\end{equation}
	where $\mathbf{U}$ is the matrix of eigenvectors and $\mathbf{\Lambda} = \text{diag}(\lambda_1, \lambda_2, ..., \lambda_N)$ is the diagonal matrix of eigenvalues. 
	
	To ensure we capture most of the information, we retain only the first $k$ principal components that account for at least 95\% of the total variance. This is expressed as:
	\begin{equation}
		\sum_{i=1}^{k} \lambda_i \geq 0.95 \sum_{i=1}^{N} \lambda_i \label{2.21}
	\end{equation}
	
	The final forecasted covariance matrix is then approximated as:
	\begin{equation}
		{\widehat{\mathbf{\Sigma}}}_{t + 1:t + F} \approx \mathbf{U}_k \mathbf{\Lambda}_k \mathbf{U}_k^\top \label{2.22}
	\end{equation}
	
	\subsubsection{Random Matrix Theory ($RMT$)}
	$RMT$ \citep{Laloux2000} is built upon the $PCA$ where the upper ($\lambda_+$) and lower ($\lambda_-$) bound of the eigenvalues are calculated using the Marchenko-Pastur distribution. These bounds are calculated as:
	\begin{equation}
		\lambda_+ = \left(1 + \frac{1}{\sqrt{q}}\right)^2, \quad \lambda_- = \left(1 - \frac{1}{\sqrt{q}}\right)^2 \label{2.23}
	\end{equation}
	where $q=T/N$ is the ratio of the window size to the number of assets. Eigenvalues within these bounds are dominated by noise, while eigenvalues outside the bounds represent meaningful information.
	\begin{equation}
		\lambda_i^{\text{filtered}} = 
		\begin{cases} 
			\lambda_i & \text{if } \lambda_i > \lambda_+ \text{ or } \lambda_i < \lambda_- \\ 
			\frac{\lambda_+ + \lambda_-}{2} & \text{otherwise}
		\end{cases} \label{2.24}
	\end{equation}
	This filtering process removes noise and retains only the significant components of the covariance matrix.
	The cleaned covariance matrix is reconstructed by multiplying the filtered eigenvalues ($\mathbf{\Lambda}^{\text{filtered}}$) with their corresponding eigenvectors:
	\begin{equation}
		{\widehat{\mathbf{\Sigma}}}_{t + 1:t + F} = \mathbf{U} \mathbf{\Lambda}^{\text{filtered}} \mathbf{U}^\top \label{2.25}
	\end{equation}
	
	\subsubsection{Ledoit and Wolf ($LW$)}
	The LW shrinkage \citep{Ledoit2004} is a robust statistical model that combines the sample covariance matrix with a structured target matrix, improving forecast accuracy by reducing the impact of forecast error due to high dimensionality and limited sample size. The LW shrinkage estimator is given by:
	\begin{equation}
		{\widehat{\mathbf{\Sigma}}}_{t + 1:t + F} = \rho \mathbf{T} + (1 - \rho) \mathbf{\Sigma}_{t - F:t} \label{2.11}
	\end{equation}
	where \( \mathbf{T} = \frac{\text{Tr}(\mathbf{\Sigma}_{t - F : t})}{N} \times \mathbf{I} \) is the target matrix, with $\mathbf{I}$ being the identity matrix and $\text{Tr}(\mathbf{\Sigma}_{t - F : t})$ representing the sum of the diagonal elements of the sample covariance matrix. The shrinkage coefficient $\rho$ is optimally determined to minimise the mean squared error between the estimated and true covariance matrices:
	\begin{equation}
		\rho = \frac{\sum_{i \neq j} \operatorname{Var}(\sigma_{i,j})}{\sum_{i \neq j} (\sigma_{i,j} - t_{i,j})} \label{2.12}
	\end{equation}
	where $\sigma_{i,j}$ are the elements of the sample covariance matrix and $t_{i,j}$ are the elements of the target matrix $\mathbf{T}$.
	
	\subsubsection{Ledoit and Wolf full sample ($LW^F$)}
	Similarly to the $NA^F$ model, we employed a full sample of \citet{Ledoit2004} as a shrinkage model.
	\begin{equation}
		{\widehat{\mathbf{\Sigma}}}_{t + 1:t + F} = \rho \mathbf{T} + (1 - \rho) \mathbf{\Sigma} \label{2.13}
	\end{equation}
	
	\subsection{GARCH Models}
	\subsubsection{Constant Conditional Correlations ($CCC$)}
	The $CCC$ model, introduced by \citet{Bollerslev1990}, assumes that asset correlations remain constant over time while variances evolve dynamically according to a GARCH process based on a one-step forecast. To extend the model to a multi-step forecasting framework, we follow the methodology of \citet{Baillie1992}, defining the $F$-step ahead conditional covariance forecast as:
	\begin{equation}
		\widehat{\mathbf{\Sigma}}_{t + 1:t + F} = \frac{1}{F} \sum_{f=1}^{F} \mathbf{D}_{t+f} \mathbf{R} \mathbf{D}_{t+f}, \label{2.14}
	\end{equation}
	where $\mathbf{D}_{t+f} = \text{diag} \left( \sqrt{h_{11,t+f}}, \sqrt{h_{22,t+f}}, \dots, \sqrt{h_{NN,t+f}} \right)$ is the diagonal matrix of forecasted conditional volatilities for horizon $f$ and $R$ the conditional correlation matrix. The multi-step averaging approach used in Eq.~\eqref{2.14} was introduced by \citet{DeNard2021} to provide a more stable covariance forecast over the rebalancing period. The $h_{ii,t}$ series is modelled using univariate GARCH(1,1) processes:
	\begin{equation}
		h_{ii,t+f} = \sum_{j=0}^{f-1} \omega_i (\alpha_i + \beta_i)^j + (\alpha_i + \beta_i)^f h_{ii,t}, \label{2.15}
	\end{equation}
	where $\omega_i, \alpha_i, \beta_i$ are the estimated GARCH parameters.
	
	The conditional correlation matrix $\mathbf{R}$ is assumed to be constant over time calculated through the standardised residuals, which are defined as:
	\begin{equation}
		z_{i,t} = \frac{e_{i,t}}{\sqrt{h_{ii,t}}}, \label{2.16}
	\end{equation}
	where $e_{i,t}$ are the residuals of the return series, which are used to estimate the conditional correlation matrix.
	\subsubsection{Dynamic Conditional Correlations ($DCC$)}
	The $DCC$ model of \citet{Engle2002} extends the $CCC$ model by allowing correlations to vary over time while retaining the GARCH framework for variances. This is achieved by replacing the constant correlation matrix in Eq.~\eqref{2.14} with a dynamic correlation matrix $\mathbf{R}_t$.
	
	To accommodate multi-step forecasting for practical portfolio management, we follow the methodology of \citet{Engle2001}. The $F$-step ahead conditional covariance forecast is:
	\begin{equation}
		\widehat{\mathbf{\Sigma}}_{t + 1:t + F} = \frac{1}{F} \sum_{f=1}^{F} \mathbf{D}_{t+f} \mathbf{R}_{t+f} \mathbf{D}_{t+f}, \label{2.17}
	\end{equation}
	where $\mathbf{D}_{t+f}$ is defined as in Eq.~\eqref{2.14}, and $\mathbf{R}_{t+f}$ is the $f$-step ahead forecast of the dynamic conditional correlation matrix. The multi-step forecast for the correlation matrix is computed as follows:
	\begin{equation}
		\mathbf{R}_{t+f} = \sum_{j=0}^{f-1}(1 - \alpha - \beta) \overline{\mathbf{Q}} (\alpha + \beta)^j + (\alpha + \beta)^f \mathbf{R}_t, \label{2.18}
	\end{equation}
	where $\overline{\mathbf{Q}}$ is the unconditional correlation matrix, and $\alpha + \beta < 1$ ensures stationarity. The parameters $\alpha$ and $\beta$ control the persistence of the dynamic correlations.
	
	The conditional correlation matrix $\mathbf{R}_t$ evolves over time according to:
	\begin{equation}
		\mathbf{R}_{t} = \mathbf{V}_{t}^{-1} \mathbf{Q}_{t} \mathbf{V}_{t}^{-1} \label{2.19}
	\end{equation}
	\begin{equation}
		\mathbf{Q}_{t} = (1 - \alpha - \beta) \overline{\mathbf{Q}} + \alpha \mathbf{z}_{t-1} \mathbf{z}_{t-1}' + \beta \mathbf{Q}_{t-1}, \label{2.20}
	\end{equation}
	where $\mathbf{V}_{t} = \text{diag}\{\sqrt{q_{11,t}}, \sqrt{q_{22,t}}, \dots, \sqrt{q_{NN,t}}\}$ ensures that $\mathbf{R}_t$ remains a proper correlation matrix. The standardised residuals $\mathbf{z}_t$ are defined as in Eq.~\eqref{2.16}, and $\overline{\mathbf{Q}}$ is estimated from historical standardised residuals.
	
	\subsubsection{Nonlinear Shrinkage Dynamic Conditional Correlations ($DCC^{NL}$)}
	
	One work that combines the multivariate GARCH with a nonlinear shrinkage is the model of \citet{Engleetal2019}. The covariance forecast in the $DCC^{NL}$ model follows the same structure as in Eq.~\eqref{2.17}, but now with a shrinkage-improved correlation matrix ($\overline{\mathbf{Q}}$). The original model uses the numerical method of \citet{Ledoit2015}. However, we apply the nonlinear shrinkage method of \citet{Ledoit2020} to speed up the computations, which optimally shrinks the eigenvalues of the sample covariance matrix towards a better-conditioned target matrix. 
	
	Given a sample correlation matrix $\mathbf{\Upsilon}$, the optimal nonlinear shrinkage estimator adjusts its eigenvalues as follows. Let $\mathbf{\Upsilon}$ have the eigendecomposition:
	\begin{equation}
		\mathbf{\Upsilon} = \mathbf{U} \mathbf{\Lambda} \mathbf{U}^{\top},
	\end{equation}
	The nonlinear shrinkage transformation replaces $\mathbf{\Lambda}$ with a shrinkage-transformed version $\mathbf{\Lambda}^{*}$:
	\begin{equation}
		\mathbf{\Upsilon}^{*} = \mathbf{U} \mathbf{\Lambda}^{*} \mathbf{U}^{\top} \label{2.22}
	\end{equation}
	where the optimally shrunk eigenvalues $\lambda_i^{*}$ are computed using the oracle nonlinear shrinkage function:
	\begin{equation}
		\lambda_i^{*} = \lambda_i \frac{1}{\left( \pi c \lambda_i f(\lambda_i) \right)^2 + \left( 1 - c - \pi c \lambda_i H_f(\lambda_i) \right)^2}
	\end{equation}
	where \( f(\lambda_i) \) is the sample spectral density estimator, \( H_f(\lambda_i) \) is the Hilbert transform of the spectral density, and \( c = \frac{N}{T} \) is the limiting concentration ratio of the covariance matrix.
	
	The sample spectral density function \( f(x) \) is estimated using a kernel-based density estimator, ensuring uniform consistency. The Hilbert transform \( H_f(x) \) is computed using the principal value integral, which corrects for eigenvalue bias induced by sampling variability.
	
	The shrunk correlation matrix ($\mathbf{\Upsilon}^{*}$) is then used as the input for the $DCC$ process in Eq.~\eqref{2.19}. 
	
	\subsection{CNN with A-BiLSTM ($CAB$)}
	In our proposed model, each batch of input data is processed sequentially through all six stages, as described below.
	
	\subsubsection{Data Preprocessing}
	\label{datapre}
	Let $\mathbf{B} = [\mathbf{D}_1, \mathbf{D}_2, \dots, \mathbf{D}_T]^\top $ be the time series data, where each $\mathbf{D}_t \in \mathbb{R}^{N \times N} $ represents the rolling covariance matrix, normalised, of $N$ assets at time $t$). 
	
	It is essential to standardise the input features to facilitate efficient training and convergence of the $CAB$ model. This transformation ensures that each feature has zero mean and unit variance, which is crucial for stabilising the training dynamics of neural networks.

	We transform the data into overlapping sequences using a rolling window approach to capture temporal dependencies. Specifically, for a given lookback period $L$, the sequence creation process, at time $t$, can be expressed as:
	\begin{equation}
		\mathbf{S}_t = [\mathbf{D}_{t-L}, \mathbf{D}_{t-L+1}, \dots, \mathbf{D}_{t}]^\top \in \mathbb{R}^{L \times N \times N} \label{2.27}
	\end{equation}
	
	The input to the $CAB$ model consists of sequences of scaled covariance matrices. Formally, let \( \mathbf{S}_i \in \mathbb{R}^{L \times N \times N} \) denote the \( i \)-th sequence.	
	
	\subsubsection{3D Convolutional Layer}
	Next, each sequence of covariance matrices is processed through a three-dimensional layer \citep{LeCun1998}. We use a 3D CNN to capture local spatio-temporal features, given that each rolling covariance matrix is an ($N \times N$) grid, and the temporal dimension arises from stacking these grids over time. The 3D CNN filters can identify how subsets of assets interact and evolve over multiple past windows, offering a more flexible approach than linear assumptions.
	
	By applying a convolution operation with a kernel size of (\( ks \times ks \times ks \)), the same padding, and a stride equal to one, the CNN layer captures patterns within the data across different time steps and different dimensions of the covariance matrices. The transformation performed by the 3D convolutional layer on the input sequence of covariance matrices can be expressed as:
	
	\begin{equation}
		\mathbf{X}_{\text{conv}}(t, i, j) = \sum_{u=-\lfloor ks/2 \rfloor}^{\lfloor ks/2 \rfloor} \sum_{v=-\lfloor k_s/2 \rfloor}^{\lfloor ks/2 \rfloor} \sum_{w=-\lfloor ks/2 \rfloor}^{\lfloor ks/2 \rfloor} \mathbf{W}(u, v, w) \cdot \mathbf{X}_{\text{rs}}(t+u, i+v, j+w) + b \label{2.30}
	\end{equation}
	where $ks$ is the kernel size of the 3D Convolution, determining the dimensions of the receptive field across the temporal and spatial axes. $\mathbf{W}(u, v, w)$ represents the kernel weights, a learnable parameter of size $C_{\text{out}} \times C_{\text{in}} \times k_s \times k_s \times k_s$, where $C_{\text{in}}$ is the number of input channels, and $C_{\text{out}}$ is the number of output channels. The input covariance matrices are reshaped to include a single channel, so $C_{\text{in}} = 1$. Additionally, the model outputs a single channel, meaning $C_{\text{out}} = 1$. Consequently, the weight tensor simplifies to $\mathbf{W} \in \mathbb{R}^{1 \times 1 \times ks \times ks \times ks}$. $\mathbf{X}_{\text{rs}}(t+u, i+v, j+w)$ is the input sequence of covariance matrices reshaped to dimensions $1 \times L \times N \times N$. The term $b$ denotes the bias added to the convolution result, which, with $C_{\text{out}} = 1$, is a single scalar, i.e., $b \in \mathbb{R}^{1}$. The indices $u$, $v$, and $w$ iterate over all whole integers within the range $[-( k_s / 2 ), ( k_s / 2 )]$, centered around $(t, i, j)$, to compute local features. Finally, zero-padding is applied to ensure that the output dimensions match the input dimensions, preserving the temporal and spatial resolution of the covariance matrices.

	Subsequently, the convolutional output is flattened ($\mathbf{X}_{\text{flattened}}\in \mathbb{R}^{L \times N^2}$) to transition from spatial to feature dimensions suitable for the LSTM layer. This reshaping consolidates the \( N \times N \) covariance matrices into \( N^2 \)-dimensional feature vectors for each time step within the sequence.
	
	\subsubsection{Bidirectional LSTM (BiLSTM) Layer}
	Following the CNN, the flattened transformed sequences are fed into a layer configured with $u$ layers and $h_d$ hidden dimension. To understand the BiLSTM, it is essential first to grasp the fundamentals of a standard LSTM network \citep{Hochreiter1997}. Appendix~\ref{LSTM} provides more detailed information on how each LSTM cell works.
	
	The BiLSTM enhances the standard LSTM by adding another layer that processes the input sequence in the reverse order. This bidirectional approach allows the model to have both forward and backward information about the sequence, making it more effective in capturing the context from past and future data points. Each sequence goes through $u$ layers, i.e. the output of layer one serves as input for layer 2.
	
	In our model, the BiLSTM first transforms the input sequence $\mathbf{X}_{\text{flattened}}$ of dimension $L \times N^2$ into an output sequence of dimension $L \times 2h_d$. The BiLSTM uses learnable parameters to map the $N^2$ features at each time step into a lower-dimensional hidden representation of size $h_d$. This transformation is achieved through linear transformations, where weights and biases are optimised during training. 	
	Then, the hidden states from both forward and backward pass across all layers, and time steps are concatenated to form a comprehensive hidden state matrix. This matrix captures information from all directions and spans the entire sequence of inputs. Each BiLSTM layer can be expressed as:
	
	\begin{equation}
		\mathbf{H}^o = 
		\begin{bmatrix}
			\mathbf{H}^{(1)}, \mathbf{H}^{(2)} , \dots, \mathbf{H}^{(h_d)} \\
		\end{bmatrix} \in \mathbb{R}^{L \times 2h_d} \label{2.38}
	\end{equation}
	where $o$ represents the BiLSTM layer, for $o = 1,\dots,u$, and each \( \mathbf{H}^{(l)} \) for \( l = 1, 2, \ldots, h_d \) is defined as:
	
	\begin{equation}
		\mathbf{H}^{(l)} = 
		\begin{bmatrix}
			\overrightarrow{\mathbf{h}}^{(l)}_{t-L} & \overleftarrow{\mathbf{h}}^{(l)}_{t-L} \\
			\overrightarrow{\mathbf{h}}^{(l)}_{t-L+1} & \overleftarrow{\mathbf{h}}^{(l)}_{t-L+1} \\
			\vdots & \vdots \\
			\overrightarrow{\mathbf{h}}^{(l)}_{t} & \overleftarrow{\mathbf{h}}^{(l)}_{t} \\
		\end{bmatrix}
		\in \mathbb{R}^{L \times 2}
		\label{2.39}
	\end{equation}
	The hidden dimension for each LSTM direction, denoted as $h_d$, results in a concatenated hidden state dimension of $2h_d$ at each time step.
	
	The traditional BiLSTM outputs the last time step of the last layer. However, in our model, we use a similar approach to \citet{Liu2019} methodology, meaning all time steps from the last layer ($\mathbf{H}^u$) are used as an input to our following steps. This procedure allows the attention mechanism to dynamically weigh the importance of each time step, enabling the model to focus on the most relevant temporal patterns when making predictions. By leveraging all time steps from the last layer ($\mathbf{H}^u$), rather than just the final output, the model can more effectively capture long-term dependencies and contextual relationships.
	
	To mitigate overfitting and enhance the model's generalisation capabilities, a Dropout Layer is applied to the BiLSTM output. This operation randomly zeroes a fraction of the elements in \( \mathbf{H}^u \) with a probability of 20\%, effectively regularising the model by preventing reliance on specific neurons during training. During our testing phase, this feature will be skipped.
	
	\subsubsection{Multi-Head Self-Attention Layer}
	Our next step is to apply a multi-head attention mechanism \citep{Vaswani2017} to each hidden state of the last layer.
	The multi-head attention mechanism to refine the temporal dependencies captured in the sequence. For $h_e$ attention heads, the input $\mathbf{H}^u$ is used to compute the queries ($\mathbf{Q}$), keys ($\mathbf{K}$), and values ($\mathbf{V}$): $\mathbf{Q} = \mathbf{H}^u \textbf{W}_q$, $\mathbf{K} = \mathbf{H}^u \textbf{W}_k$, and $\mathbf{V} = \mathbf{H}^u \textbf{W}_v$, where $\textbf{W}_q, \textbf{W}_k, \textbf{W}_v \in \mathbb{R}^{2h_d \times 2h_d}$ are learnable parameters. The attention weights for each head are computed as follows:
	\begin{equation}
		\textbf{Attention}(\mathbf{Q}_{\text{head}}, \mathbf{K}_{\text{head}}, \mathbf{V}_{\text{head}}) = \text{softmax}\left(\frac{\mathbf{Q}_{\text{head}} \mathbf{K}_{\text{head}}^\top}{\sqrt{d_k}}\right)\mathbf{V}_{\text{head}} \in \mathbb{R}^{L \times d_k} \label{2.40}	
	\end{equation}
	where \( d_k = \frac{2h_d}{h_e} \) scales the dot product to stabilise gradients and splits the features. Each head processes its respective portion of the features, which are concatenated to produce $\mathbf{A}_{concat} \in \mathbb{R}^{L \times 2h_d}$. The concatenated output is projected back to $\mathbf{A} = \mathbf{A}_{concat} \textbf{W}_c \in \mathbb{R}^{L \times 2h_d}$ using a linear transformation.
	
	The attention output is mean-pooled along the sequence dimension to condense the temporal information into a fixed-length representation:
	\begin{equation}
		\mathbf{a}_{\text{mean}} = \frac{1}{L} \sum_{t=1}^{L} \mathbf{A}_t \quad \in \mathbb{R}^{1 \times 2h_d}\label{2.41}
	\end{equation}
	where $\mathbf{A}_t$ is the $t$-th row of $\mathbf{A}$. This averaged context vector \( \mathbf{a}_{\text{mean}} \) encapsulates the overall information from the entire sequence. We then pass \( \mathbf{a}_{\text{mean}} \) through a fully connected layer to map it back to the covariance matrix space:
	
	This averaged context vector \( \mathbf{a}_{\text{mean}} \) encapsulates the overall information from the entire sequence. We then pass \( \mathbf{a}_{\text{mean}} \) through a fully connected layer to map it back to the covariance matrix space:
	
	\begin{equation}
		\mathbf{y} = \mathbf{a}_{\text{mean}} \mathbf{W}_{\text{fc}} + \mathbf{b}_{\text{fc}} \quad \in \mathbb{R}^{1 \times N^2}\label{2.42}
	\end{equation}
	where $\mathbf{W}_{\text{fc}}$ and \( \mathbf{b}_{\text{fc}} \) are receptively the weight matrix and bias vector of the fully connected layer.
	
	Because our pipeline stacks 3D Convolution, BiLSTM, and attention in succession, the final latent representation becomes highly abstract, making interpretability at each layer difficult. Consequently, we do not attempt to interpret the hidden states or attention weights in a domain sense.
	
	\subsubsection{Enforcing Symmetry and Positive Semi-Definiteness}
	The output vector \( \mathbf{y} \) is reshaped into a square matrix ($\mathbf{Y}_{\text{reshaped}} \in \mathbb{R}^{N \times N}$) to form the predicted covariance matrix.
	
	The output matrix \(\mathbf{Y}_{\text{reshaped}}\) may not necessarily be symmetric, a key property of covariance matrices. To enforce symmetry, we average the matrix with its transpose:
	\begin{equation}
		\mathbf{Y}_{\text{sym}} = \frac{\mathbf{Y}_{\text{reshaped}} + \mathbf{Y}_{\text{reshaped}}^\top}{2} \label{2.44}
	\end{equation}
	
	Then, we rescale the normalised output back to the original scale of the data, using the reverse process of Section~\ref{datapre} resulting in \(\mathbf{Y}_{\text{sym}}^{unscaled}\).
	
	Symmetry alone does not guarantee that \(\mathbf{Y}_{\text{sym}}^{unscaled}\) is positive semi-definite, which is another requisite property of covariance matrices. To ensure positive semi-definiteness, we perform eigenvalue decomposition:
	\begin{equation}
		\mathbf{Y}_{\text{sym}}^{unscaled} = \mathbf{U} \mathbf{\Lambda} \mathbf{U}^\top \label{2.45}
	\end{equation}
	We then modify \(\mathbf{\Lambda}\) by setting all negative eigenvalues to zero:
	\begin{equation}
		\mathbf{\Lambda}' = \text{diag}\left(\max(\lambda_i, 0)\right) \label{2.46}
	\end{equation}
	The positive semi-definite matrix is reconstructed as:
	\begin{equation}
		\mathbf{Y}_{\text{psd}} = \mathbf{U} \mathbf{\Lambda}' \mathbf{U}^\top \label{2.47}
	\end{equation}
	
	\subsubsection{Shrinkage}
	While DL-based forecasts can learn intricate patterns, which provide a more stable forecast, the noisy historical covariance may still convey valuable information about recent market conditions. Therefore, our model's final step is incorporating a linear shrinkage from historical values in our predictions.
	\begin{equation}
		\textbf{CAB}=\mathbf{Y}_{\text{psd}} \dot \phi+\mathbf{\Sigma}_{t - F:t} \dot (1-\phi)\label{2.48}
	\end{equation}
	where $\textbf{CAB}$ is the final $N \times N$ matrix output, and $\phi$ is the shrinkage factor. 
	
	\subsection{Evaluation metrics}
	We will use performance metrics consistent with previous literature \citep{Symitsi2018, Bollerslevetal2018, Zhang2024} to check the models' performance. We consider the following loss functions to measure the average distance between predicted covariances and realised covariances matrices for the model comparisons:
	\begin{equation}
		\mathcal{L}_t^E = \sqrt{\operatorname{vec}_{Tri}(\mathbf{\Sigma}_t-\hat{\mathbf{\Sigma}}_t ) ^\top \operatorname{vec}_{Tri} \left( \mathbf{\Sigma}_t - \mathbf{\hat\Sigma}_t \right) }\label{2.49}
	\end{equation}
	\begin{equation}
		\mathcal{L}_t^F = \sqrt{ \operatorname{Tr}\left[ \left( \bm{\Sigma}_t - \bm{\hat{\Sigma}}_t \right)^\top \left( \bm{\Sigma}_t - \bm{\hat{\Sigma}}_t \right) \right] }\label{2.50}
	\end{equation}
	where $\mathcal{L}_t^E$ represents the Euclidean distance between the $N(N+1)/2$ dimensional vectorised version of the upper triangular of forecast covariances and ex-post realised covariances. $\mathcal{L}_t^F$ is the Frobenius distance between the two matrices, and $\operatorname{Tr}$ is the trace of a square matrix. All these functions measure losses. Therefore, lower values are preferred.
	\section{Empirical Results}
	\label{sec:empirical}
	We obtain our sample data from the Refinitiv Database, covering daily adjusted closing prices of 14 major ETFs that reflect equity and bond markets across diverse global sectors. These ETFs were selected to construct a multi-asset environment that includes equity sectors (e.g. technology, financials, industrials, consumer staples) and fixed-income exposures. Table \ref{tab:info} details the ETF's information. To compute \emph{excess} returns, we also retrieve from the Federal Reserve Economic Data the 1-month U.S. Treasury yield, treating it as a daily risk-free rate. Our sample runs from 1~January~2017 to 31~December~2023, totalling 1,760 trading days.
	
	\begin{table}[h]
		\centering
		\caption{List of ETFs included in our research.}
		\label{tab:info}
		\begin{tabular}{lcccccc}
			Name&Ticker&ISIN&Asset Class\\
			\hline
			iShares Global Tech ETF&IXN&US4642872919&Equity\\
			iShares Global Financials ETF&IXG&US4642873339&Equity\\
			iShares Global Consumer Discretionary ETF&RXI&US4642887453&Equity\\
			iShares Global Industrials ETF&EXI&US4642887297&Equity\\
			iShares Global Healthcare ETF&IXJ&US4642873255&Equity\\
			iShares Global Consumer Staples ETF&KXI&US4642887370&Equity\\
			iShares U.S. Real Estate ETF&IYR&US4642877397&Real Estate\\
			iShares International Developed Real Estate ETF&IFGL&US4642884898&Real Estate\\
			iShares Global Materials ETF&MXI&US4642886950&Equity\\
			iShares Global Energy ETF&IXC&US4642873412&Equity\\
			iShares Global Comm Services ETF&IXP&US4642872752&Equity\\
			iShares Global Utilities ETF&JXI&US4642887115&Equity\\
			iShares Core U.S. Aggregate Bond ETF&AGG&US4642872265&Fixed Income\\
			iShares Core International Aggregate Bond ETF&IAGG&US46435G6724&Fixed Income\\
			
		\end{tabular}
	\end{table}
	
	We split our dataset into a training period from 1~January~2017 to 31~December~2020 and a testing period covering 1~January~2021 to 31~December~2023. The testing sample thus comprises 753 trading days, encompassing a wide range of economic and market conditions. We identify three distinct market regimes within this testing window to examine whether our model responds robustly to varying interest-rate and equity-market environments. We designate the interval between 1~January~2021 and 2~January~2022 as the \textit{First Bull Period} (Bull-1), typified by low yields on the risk-free asset and a generally stable equity market. We label the interval from 3~January~2022 to 12~June~2022 as the \textit{Bear Period} (Bear), characterised by a steady rise in the risk-free rate and a drop of 20\% or more in the S\&P~500 from its last peak. Lastly, we classify the interval from 13~June~2022 to 31~December~2023 as the \textit{Second Bull Period} (Bull-2), featuring persistently high interest rates but a renewed upswing in equity prices. This subdivision ensures that our out-of-sample evaluations capture both low- and high-rate market phases and a bear-market regime, offering a comprehensive test of the model's ability to adapt to evolving financial landscapes.
	
	During the training period, we minimise the Frobenius distance (Eq.~\ref{2.50}) to select hyperparameters. We use \textit{Optuna} \citep{Akiba2019} for hyperparameter selection. Appendix~\ref{hyper} shows the grid search table. Specifically, we split the training dataset into a model-fitting subset (80\%) and a validation subset (20\%) to identify optimal settings under a forecast horizon ($F$) of 20 days. We adopt the Adam optimiser over 100 epochs, using a batch size of 128 and a learning rate of $10^{-4}$. For the neural-network-based $CAB$ model, we use a lookback window ($L$) of 100 days. We set the 3D convolution kernel size ($ks$) to five; the BiLSTM has 128 hidden dimensions ($hd$) across seven stacked layers ($u$), and the multi-head attention uses 16 heads ($he$). Finally, we apply a shrinkage factor of 0.8 ($\phi$) when combining the DL output with the most recent historical covariance. 
	
	We conduct a rolling forecast procedure to simulate live conditions. Each new trading day $t$ in the test set triggers an online update of our DL model, incorporating all data up through day $t$. This design ensures the forecasts mimic a real-world trading environment, where only up-to-date historical data is available at each decision point.
	
	Table \ref{tab:OResults} reports the average Euclidean ($\mathcal{L}^E$) and Frobenius ($\mathcal{L}^F$) distances between forecasted and realised covariance matrices for all models over the entire testing period (2021--2023)\footnote{Bold and underlined numbers represent the lowest and the second lowest values, respectively.}. GARCH models ($CCC$ and $DCC$) and $EWMA$ perform relatively well as they adapt faster to structural changes in the covariance matrices. Please note that although $DCC^{NL}$ is not more accurate than the traditional $DCC$, it provides more stable forecasts based on the eigenvalues. However, our model ($CAB$) yields the lowest errors in both metrics, outperforming classical benchmarks. Dimensional reduction models perform inadequately as we only use 14 assets.
	
	\begin{table}[h]
		\centering
		\caption{Average Euclidean ($\mathcal{L}^E$) and Frobenius ($\mathcal{L}^F$) distances ($\times 10^5$) for whole the testing period (1~January~2021 to 31~December~2023).}
		\label{tab:OResults}
		\renewcommand{\arraystretch}{1.15}
		\begin{tabular}{lrr}
			& \multicolumn{1}{c}{$\mathcal{L}^E$} & \multicolumn{1}{c}{$\mathcal{L}^F$} \\
			\cline{2-3}
			$NA$ &53.526&82.020\\
			$NA^F$&52.713&80.987\\
			$EWMA$&49.111&75.349\\
			$PCA$ &54.348&83.154\\
			$RMT$&54.145&82.915\\
			$LW$ &49.772&76.514\\
			$LW^F$&68.136&103.363\\
			$CCC$ &48.568&75.198\\
			$DCC$ &\underline{48.268}&\underline{74.716}\\
			$DCC^{NL}$&51.621&79.079\\
			$CAB$&\textbf{38.312}&\textbf{58.923}\\
			
		\end{tabular}
	\end{table}
	
	To solidify our findings, we analysed the 10 models with 753 paired samples for the Euclidean and Frobenius distances following \citet{Demsar2006} frequentist methodology. The significance level is $\alpha=0.050$ for all tests here forward. We rejected the null hypothesis that the population is normal for all the populations ($p < 0.01$ for all models in both distances). Therefore, we assume that not all populations are normal.
	
	Because we have more than two populations and the populations are not normal, we use the non-parametric Friedman test as an omnibus test to determine if there are any significant differences between the median values of the populations. We use the post-hoc Nemenyi test to infer which differences are significant. Differences between populations are significant if the difference in the mean rank is greater than the critical distance CD=0.494 of the Nemenyi test.
	
\begin{figure}[htbp]
		\centering
		\begin{subcaption}{Upper Triangular Euclidean distance}
			\centering
			\includegraphics[width=\linewidth]{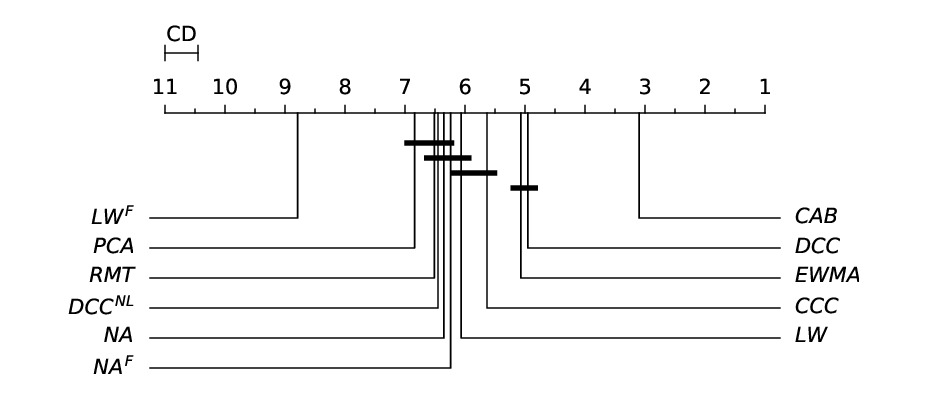}
		\end{subcaption}
		\begin{subcaption}{Frobenius distance}
			\centering
			\includegraphics[width=\linewidth]{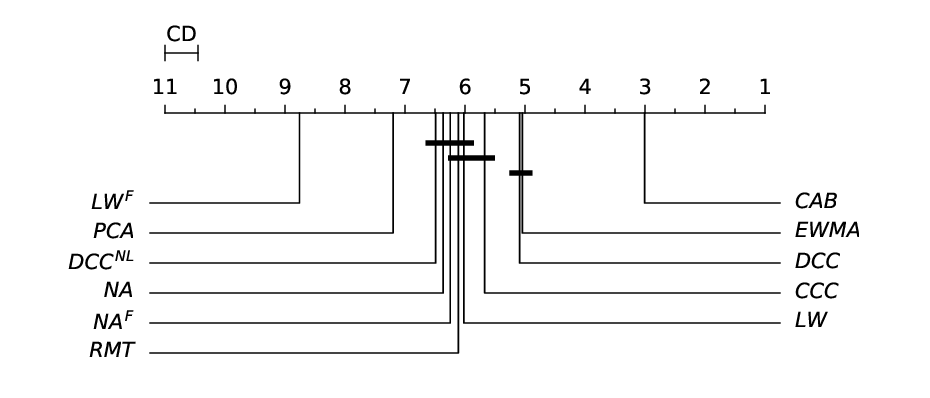}
		\end{subcaption}
		\caption{Critical distance diagrams for the post-hoc Nemenyi test.}
		\label{fig:combined_plot}
	\end{figure}

	We reject the Friedman test's null hypothesis, which states that there is no difference in the central tendency of the populations. Therefore, we assume that there is a statistically significant difference between the central tendencies of the populations. 
	Figure~\ref{fig:combined_plot} shows the post-hoc Nemenyi test critical distance between groups. Our model reports statistically better results than the benchmark models.
	
	Table \ref{tab:periods} presents the average distance metrics for the testing period across three market regimes: Bull-1, Bear, and Bull-2. Overall, the $CAB$ model consistently outperforms all benchmarks.
	
	\begin{table}[h]
		\centering
		\caption{Average Euclidean ($\mathcal{L}^E$) and Frobenius ($\mathcal{L}^F$) distances ($\times 10^5$) for different testing periods: Bull-1 (1~January~2021 to 2~January~2022), Bear (3~January~2022 to 12~June~2022), and Bull-2 (13~June~2022 to 31~December~2023).}
		\label{tab:periods}
		\begin{tabular}{lcclcclclcc}
			& \multicolumn{2}{c}{Bull-1} && \multicolumn{2}{c}{Bear} & & \multicolumn{2}{c}{Bull-2} \\
			& $\mathcal{L}^E$ & $\mathcal{L}^F$ & &$\mathcal{L}^E$ & $\mathcal{L}^F$ & & $\mathcal{L}^E$ & $\mathcal{L}^F$ \\
			\cline{2-3}
			\cline{5-6}
			\cline{8-9}			
			$NA$ &36.937&56.621&&89.891&137.868&&53.709&82.252\\
			$NA^F$&35.680&54.992&&98.002&150.780&&50.586&77.545\\
			$EWMA$&31.754&\underline{48.973}&&\underline{82.799}&\underline{126.614}&&50.574&77.549\\
			$PCA$ &37.424&57.281&&90.720&139.116&&54.747&83.660\\
			$RMT$&37.246&57.083&&90.638&138.821&&54.493&83.411\\
			$LW$ &34.126&52.599&&90.441&138.173&&48.089&74.088\\
			$LW^F$&63.288&94.759&&93.189&143.636&&63.989&97.222\\
			$CCC$ &\underline{31.688}&49.784&&88.106&134.902&&48.017&74.316\\
			$DCC$ &33.516&52.166&&84.745&130.345&&\underline{47.227}&\underline{73.161}\\
			$DCC^{NL}$&36.779&56.449&&87.650&134.167&&50.767&77.734\\
			$CAB$&\textbf{24.276}&\textbf{37.643}&&\textbf{66.381}&\textbf{102.252}&&\textbf{39.253}&\textbf{60.126}\\
			
		\end{tabular}
	\end{table}
	
	While all models experience a decrease in accuracy during the turbulent Bear regime, the $CAB$ approach exhibits notable robustness. In particular, during the Bull-2 period, characterised by elevated interest rates and dynamic shifts in cross-asset relationships, integrating DL techniques with classical shrinkage enables $CAB$ to adapt effectively to changing market conditions, thereby maintaining lower distance metrics.
	
	Supplementary Nemenyi post-hoc tests confirm that the performance improvements observed with $CAB$ are statistically significant. These tests provide further evidence that the model's superior performance is not merely a result of sample-specific idiosyncrasies but reflects a robust ability to handle varying market regimes.
	
	Moreover, these findings contribute to the literature on hybrid forecasting methods by demonstrating that blending modern DL with traditional econometric techniques can offer substantial gains in predictive accuracy, especially in periods of market stress. This robust performance under diverse macroeconomic conditions highlights the practical relevance of the $CAB$ approach for risk management and strategic asset allocation.
	
	\section{Robustness analysis}
	\label{sec:robust}
	\subsection{Larger dataset}
	\label{subsec:larger}
	In this section, we expand our asset universe to 260 ETFs to evaluate whether our model continues to outperform established benchmarks in a high-dimensional setting. Appendix~\ref{tab:listofetfslarge} provides a comprehensive list of the ETFs in a multi-asset portfolio. Table~\ref{tab:large} presents the corresponding performance metrics.
	
	\begin{table}[h]
			\centering
			\caption{Average Euclidean ($\mathcal{L}^E$) and Frobenius ($\mathcal{L}^F$) distances ($\times 10^3$) using 260 ETFs for $F=20$ and \textit{excess} returns  for different testing periods: Overall (1~January~2021 to 31~December~2023), Bull-1 (1~January~2021 to 2~January~2022), Bear (3~January~2022 to 12~June~2022), and Bull-2 (13~June~2022 to 31~December~2023).}
			\small
			\renewcommand{\arraystretch}{1.15}
			\label{tab:large}
			\begin{tabular}{lccccccccccc}
					&\multicolumn{2}{c}{Overall}&& \multicolumn{2}{c}{Bull-1} && \multicolumn{2}{c}{Bear} & & \multicolumn{2}{c}{Bull-2}\\
					& $\mathcal{L}^E$ & $\mathcal{L}^F$ & & $\mathcal{L}^E$ & $\mathcal{L}^F$ & &$\mathcal{L}^E$ & $\mathcal{L}^F$ & & $\mathcal{L}^E$ & $\mathcal{L}^F$ \\
					\cline{2-3}
					\cline{5-6}
					\cline{8-9}
					\cline{11-12}
					$NA$ &11.546&16.484&&8.677 &12.442 &&20.458&29.257&&10.814&15.390\\
					$NA^F$&10.792&15.399&&7.661&11.014 &&20.742&29.532&&9.927&14.130\\
					$EWMA$&10.544&15.054&&7.518&10.802&&18.410&26.292&&10.219&14.543\\
					$LW$ &10.719&15.304&&8.095&11.619 &&19.933&28.458&&9.740&13.867\\
					$LW^F$&13.460&19.170&&12.098&17.270 &&19.374&27.609&&12.622&17.947\\
					$CCC$&\underline{10.260}&\underline{14.647}&&\underline{7.368}&\underline{10.587} &&18.723&26.687&&9.674&13.777\\
					$DCC$&10.280&14.671&&7.796&11.182&&\underline{18.076}&\underline{25.777}&&\underline{9.622}&\underline{13.703}\\
					$DCC^{NL}$&11.047&15.747&&8.597&12.301&&18.687&26.634&&10.414&14.814\\
					$PCA$&11.658&16.643&&8.743&12.535 &&20.625&29.500&&10.939&15.567\\
					$RMT$&11.652&16.636&&8.738&12.529 &&20.618&29.490&&10.934&15.561\\
					$CAB$&\textbf{8.933}&\textbf{12.789}&&\textbf{5.447}&\textbf{7.893} &&\textbf{15.722}&\textbf{22.456}&&\textbf{9.220}&\textbf{13.153}\\
				\end{tabular}
	\end{table}
	
	Our proposed approach ($CAB$) consistently outperforms the benchmark models, reflected by its lower average Euclidean and Frobenius distances across most periods. Notably, during the Overall, Bull-1, and Bull-2 periods, the post-hoc Nemenyi test confirms that the outperformance of CAB is statistically significant. In contrast, during the Bear period, CAB’s performance is statistically similar to that of the $DCC$ and $EWMA$ models.
			
	\subsection{Impact of Using Raw Returns vs. Excess Returns}
	
	One potential criticism of our model is that it may be overly tailored to \emph{excess} returns, even though some classic optimisation frameworks \citep{Markowitz1952} often use \emph{raw} returns. We re-estimate our model without subtracting the risk-free rate to address this concern. All other model parameters remain identical to those described in Section~\ref{sec:methodology}, with the sole difference being that $\mathbf{r}_t$ now represents the \(N\)-dimensional row vector of \emph{raw} daily returns for \(N\) assets at time \(t\).
	
	Table~\ref{tab:rawreturns} reports the performance of the various models under these \emph{raw} returns assumptions. Our approach ($CAB$) continues outperforming all benchmarks across the entire sample and within individual market regimes, with statistically significant results based on the same post-hoc Nemenyi criteria used earlier. This finding directly informs our second research question about whether \emph{raw} vs.\ \emph{excess} returns lead to materially different covariance forecasts. When comparing Table~\ref{tab:rawreturns} (\emph{raw} returns) to Tables~\ref{tab:OResults} and~\ref{tab:periods} (\emph{excess} returns), the numerical magnitudes change slightly, but the rank order remains the same: $CAB$ still produces the most accurate forecasts overall.
	
	\begin{table}[h]
		\centering
		\caption{Average Euclidean ($\mathcal{L}^E$) and Frobenius ($\mathcal{L}^F$) distances ($\times 10^5$) using \emph{raw} returns for different testing periods: Overall (1~January~2021 to 31~December~2023), Bull-1 (1~January~2021 to 2~January~2022), Bear (3~January~2022 to 12~June~2022), and Bull-2 (13~June~2022 to 31~December~2023).}
		\small
		\renewcommand{\arraystretch}{1.15}
		\label{tab:rawreturns}
		\begin{tabular}{lrrrrrrrrrrr}
			&\multicolumn{2}{c}{Overall}&& \multicolumn{2}{c}{Bull-1} && \multicolumn{2}{c}{Bear} & & \multicolumn{2}{c}{Bull-2}\\
			& \multicolumn{1}{c}{$\mathcal{L}^E$}& \multicolumn{1}{c}{$\mathcal{L}^F$} & & \multicolumn{1}{c}{$\mathcal{L}^E$}& \multicolumn{1}{c}{$\mathcal{L}^F$} & & \multicolumn{1}{c}{$\mathcal{L}^E$}& \multicolumn{1}{c}{$\mathcal{L}^F$} && \multicolumn{1}{c}{$\mathcal{L}^E$}& \multicolumn{1}{c}{$\mathcal{L}^F$} \\
			\cline{2-3}
			\cline{5-6}
			\cline{8-9}
			\cline{11-12}
			$NA$ &53.525&82.020&&36.936&56.620&&89.895&137.875&&53.707&82.249\\
			$NA^F$&52.486&80.668&&34.913&53.883&&98.623&151.717&&50.462&77.372\\
			$EWMA$&49.174&75.441&&31.753&\underline{48.972}&&\underline{82.798}&\underline{126.613}&&50.697&77.730\\
			$PCA$ &54.348&83.153&&37.423&57.280&&90.724&139.123&&54.745&83.657\\
			$RMT$ &54.144&82.914&&37.245&57.081&&90.643&138.829&&54.491&83.408\\
			$LW$ &49.771&76.512&&34.125&52.598&&90.442&138.174&&48.087&74.085\\
			$LW^F$&68.130&103.353&&63.273&94.736&&93.190&143.636&&63.985&97.217\\
			$CCC$ &48.508&75.106&&\underline{31.599}&49.634&&88.139&135.038&&47.948&74.196\\
			$DCC$ &\underline{48.232}&\underline{74.661}&&33.445&52.045&&84.884&130.630&&\underline{47.161}&\underline{73.050}\\
			$DCC^{NL}$&51.604&79.051&&36.691&56.305&&87.856&134.535&&50.731&77.664\\
			$CAB$ &\textbf{36.663}&\textbf{56.518}&&\textbf{22.807}&\textbf{35.673}&&\textbf{62.036}&\textbf{96.406}&&\textbf{38.273}&\textbf{58.441}\\
			
		\end{tabular}
	\end{table}
	
	Our findings indicate that the advantage of our framework is not contingent on the choice between \emph{excess} or \emph{raw} returns, reaffirming its robustness to the presence or absence of the risk-free rate. Moreover, this consistency suggests that subtracting the risk-free rate over the medium-term horizon does not significantly alter the core structure of asset co-movements. Hence, even though economic theory posits that \emph{excess} returns better capture risk premiums, our empirical results show that the forecasting edge of $CAB$ holds across both \emph{raw} and \emph{excess} return specifications. In other words, our method retains its superior performance even when modelling raw returns, suggesting that its effectiveness does not hinge on subtracting the risk-free rate.
	
	\subsection{Varying Forecast Horizons}
	
	We also examine our model's behaviour as the forecast horizon ($F$) varies. In the baseline analysis, $F=20$ days are medium-term, but many institutional investors face different time horizons for risk management or asset allocation decisions. In line with our first research question regarding underexplored medium-term horizons, we extend $F$ from 10 to 250 days, thus covering short, medium, and long horizons. To explore model performance across this range of settings, we evaluate both \emph{excess} (Table~\ref{tab:diffhorizonsexc}) and \emph{raw} returns (Table~\ref{tab:diffhorizonsraw}).
	
	Table~\ref{tab:diffhorizonsexc} shows that our model consistently achieves the lowest average distance metrics across most forecast horizons. $CAB$ remains among the top performers for shorter horizons ($F=10$). However, it does not outperform GARCH-based models ($CCC$ and $DCC$) and $EWMA$. These differences are not statistically different in the overall period, Bull-1 and Bull-2. During Bull-2, our model outperformed the benchmarks, and the results were statistically significant based on the post-hoc Nemenyi test.
	
	For mid-term forecasting ($F=40,60,90,120,180$) for both distances, the differences are statistically significant in all periods. Thus, our model consistently outperforms the benchmarks. Please note that GARCH models tend to underperform as the forecast horizon increases, while full data models ($NA^F$ and $LW^F$) tend to diminish forecasting errors. Please note, while the $DCC^{NL}$ was more recently designed for high dimensional matrices, it provides more stable forecasts based on eigenvalues than the classical $DCC$. 
	
	At the longest horizon ($F=250$), $CAB$ again excels in most periods. Besides showing a large difference between distances during the Bear period and Bull-1, the post-hoc Nemenyi test shows no statistical difference between our model and $NA^F$ and $LW^F$, respectively. These results suggest that more complex models capture slower-moving cross-asset dynamics effectively, while classical estimators remain competitive in certain regime-specific conditions. As anticipated, naive and shrinkage methods perform relatively well at the longer horizon. These patterns reinforce our first research question's premise that medium-term horizons can benefit from sophisticated models and highlight instances where simpler approaches hold their own.
	
	Turning to Table~\ref{tab:diffhorizonsraw}, we link our findings to the second research question: whether using \emph{raw} or \emph{excess} returns makes a systematic difference.
	Results show that the top-performing models remain consistent across all forecast horizons.
	
	Once again, $CAB$ tends to yield the strongest overall performance. Although it does not dominate at shorter horizons, it consistently ranks near the top, showing no statistical difference between our model and the top performers. All other differences are statistically significant. These results mirror the trends seen under \emph{excess} returns, indicating that the performance edge of $CAB$ does not hinge on subtracting the risk-free rate. Therefore, our DL approach remains robust even if practitioners opt for \emph{raw} returns, common in certain classical optimisation frameworks. Our model performs better than the benchmarks for longer horizons ($F=250$). However, the results are not statistically different from the second-performing models ($NA^F$ and $LW^F$) for the Bull-1 and Bear periods, respectively.
	
	These findings indicate that DL-based models can adapt effectively across different forecast windows. Across both dimensions, $CAB$ shows a stable advantage, with only minor variations in extreme horizons.
	
	\section{Economic Value in Portfolio Management}
	\label{sec:portfolio}
	
	Next, we investigate the economic importance of accurate covariance forecasts in a practical asset allocation framework. Specifically, we consider an investor allocating wealth among the ETFs specified in Section~\ref{sec:empirical}, subject to a no-short-selling constraint. Such a constraint is common among regulated institutional investors (e.g., pension funds, mutual funds) that may face limits on leverage or short positions. We adopt daily, weekly, and monthly rebalancing intervals \citep{Symitsi2018}.
	
	At each rebalancing date, the investor solves the GMV problem:
	
	\begin{equation}
		\min_{\bm{w}_t} \quad \mathbf{w}_t^\top \,\widehat{\mathbf{\Sigma}}_t \,\mathbf{w}_t 
		\quad \text{subject to} \quad 
		\mathbf{1}^\top \mathbf{w}_t = 1, 
		\quad w_{i, t} \ge 0,
		\label{2.51}
	\end{equation}
	
	where $\mathbf{w}_t$ is an $N \times 1$ vector of GMV portfolio weights, $\widehat{\mathbf{\Sigma}}_t$ is the $N \times N$ forecast covariance from each model, and $\mathbf{1}$ is an $N \times 1$ vector of ones. Excluding expected returns from the optimisation allows us to isolate the impact of covariance forecasts \citep{DeMiguel2009, Kourtis2012}.
	
	Using a rolling forecast window, we generate 11 portfolio strategies (one for each forecast model) plus an equally weighted (1/$N$) benchmark. After estimating $\widehat{\mathbf{\Sigma}}_t$, we compute the GMV weights and hold them until the next rebalancing date. We then record the ex-post average portfolio return for each model ($m$) as:
	
	\begin{equation}
		\mathbf{r}^{(m)}_{t+1}
		= \frac{1}{P} \sum_{t=1}^{P} \mathbf{w}^{(m)}_{t}{}^\top \,\mathbf{r}_{t+1}, 
		\label{2.52}
	\end{equation}
	
	where $\mathbf{r}_t$ is an $N \times 1$ vector of realised asset returns, $P$ is the total out-of-sample length, and $\mathbf{w}^{(m)}_{t}$ are the portfolio weights at the start of period $t$ for model $m$.
	
	We focus on two main metrics to compare forecast models: out-of-sample portfolio \emph{variance} and \emph{turnover}. The variance of daily returns by a portfolio constructed by model $m$ is:
	
	\begin{equation}
		\sigma_m^2 
		= \frac{1}{P} \sum_{t=1}^{P} \bigl(r^{(m)}_t - \bar{r}^{(m)}\bigr)^2
		\label{2.53}
	\end{equation}
	
	where $r^{(m)}_t$ is the ex-post return of model $m$ in period $t$, $\bar{r}^{(m)}$ and $P$ is its mean and number of observations over the out-of-sample window. Meanwhile, the average turnover reflects how often the portfolio adjusts its weights:
	
	\begin{equation}
		\text{TO}_m
		= \frac{1}{P-1}
		\sum_{t=1}^{P-1} 
		\bigl\lVert 
		\mathbf{w}_{t+1}^{(m)} - \mathbf{w}_{t+}^{(m)}
		\bigr\rVert_1,
		\label{2.54}
	\end{equation}
	
	where $\mathbf{w}_{t+}^{(m)}$ are the portfolio weights of model $m$ immediately before rebalancing at $t+1$, and $\lVert \cdot \rVert_1$ is the 1-norm. A higher turnover implies more frequent portfolio adjustments, potentially incurring greater trading costs \citep{Kourtis2014}.
	
	We also include an equally weighted (1/$N$) portfolio, often cited for its simplicity and historically strong performance \citep{DeMiguel2009}. Although $1/N$ avoids estimation risk and typically maintains low turnover, it cannot adapt to shifting market correlations.
	
	From Table~\ref{tab:portfoliovariances}, it is evident that the covariance forecasts provided by the $CAB$ model strike a competitive balance between risk reduction and turnover across all rebalancing horizons. Across daily, weekly, and monthly rebalancing horizons, $CAB$ slightly trails the top daily performer, approaches the leading weekly method, and matches or surpasses most other approaches monthly, thus consistently ranking among the best-performing models. This robust performance does not come at the cost of excessive turnover, as CAB's turnover measures remain moderate compared to methods that may obtain marginally lower variances only at the expense of frequent portfolio rebalancing (e.g., $NA$, $DCC$, $DCC^{NL}$). 
	
	\begin{table}[h]
		\centering
		\caption{performance of the global minimum variance portfolios constructed using the covariance forecasts from the 11 models under consideration plus the equal-weighted portfolio. The portfolios are compared based on their annualised out-of-sample standard deviation ($\sigma_m$) and average out-of-sample turnover ($\text{TO}_m$), respectively. Results for three different rebalancing frequencies are presented.}
		\label{tab:portfoliovariances}
		\begin{tabular}{lllllllll}
			&\multicolumn{2}{c}{Daily} && \multicolumn{2}{c}{Weekly} && \multicolumn{2}{c}{Monthly}\\
			& \multicolumn{1}{c}{$\sigma_m^2$} & \multicolumn{1}{c}{$TO_m$}&& \multicolumn{1}{c}{$\sigma_m^2$} & \multicolumn{1}{c}{$TO_m$}&& \multicolumn{1}{c}{$\sigma_m^2$} & \multicolumn{1}{c}{$TO_m$}\\
			\cline{2-3}
			\cline{5-6}
			\cline{8-9}
			$1/N$&0.018132&0.005733&&0.018115&0.002567&&0.018060&0.001282\\
			$NA$&\textbf{0.001991}&0.125568&&0.002527&0.055395&&0.002631&0.025707\\
			$NA^F$&0.002451&\textbf{0.001445}&&0.002453&\textbf{0.000653}&&\textbf{0.002461}&\textbf{0.000369}\\
			$EWMA$&0.002475&0.079388&&0.002482&0.037928&&0.002554&0.016088\\
			$PCA$&0.002680&0.197322&&0.002722&0.072903&&0.002657&0.023108\\
			$RMT$&0.002647&0.125512&&0.002689&0.055954&&0.002733&0.024767\\
			$LW$&0.002655&0.098020&&0.003373&0.047089&&0.003598&0.022957\\
			$LW^F$&0.002774&\underline{0.001816}&&0.002777&\underline{0.000956}&&0.002784&\underline{0.000620}\\
			$CCC$&0.002281&0.126519&&0.002580&0.049839&&0.002713&0.014816\\
			$DCC$&0.002075&0.116446&&\textbf{0.002388}&0.051710&&\underline{0.002524}&0.017159\\
			$DCC^{NL}$&0.002048&0.129453&&0.002448&0.056173&&0.002618&0.023470\\
			$CAB$&\underline{0.002293}&0.078799&&\underline{0.002428}&0.048383&&0.002579&0.023807\\
		\end{tabular}
	\end{table}
	
	When compared to the $1/N$ benchmark, please note that all models present a statistically significant lower annual variance, based on the one-side F-Test of equalities of variance.
	\\
	
	\begingroup
	\begin{longtable}{llllllllll}
		\caption{Performance of the GMV portfolios constructed using the covariance forecasts from the 11 models under consideration plus the equal-weighted portfolio for the different market regimes. The portfolios are compared based on their annualised out-of-sample standard deviation ($\sigma_m$) and average out-of-sample turnover ($\text{TO}_m$), respectively.}\label{tab:portfoliovariancesmr}\\
		&&\multicolumn{2}{c}{Daily} && \multicolumn{2}{c}{Weekly} && \multicolumn{2}{c}{Monthly}\\
		&& \multicolumn{1}{c}{$\sigma_m^2$} & \multicolumn{1}{c}{$TO_m$}&& \multicolumn{1}{c}{$\sigma_m^2$} & \multicolumn{1}{c}{$TO_m$}&& \multicolumn{1}{c}{$\sigma_m^2$} & \multicolumn{1}{c}{$TO_m$}\\
		\cline{3-4}
		\cline{6-7}
		\cline{9-10}
		\endfirsthead 
		&&\multicolumn{2}{c}{Daily} && \multicolumn{2}{c}{Weekly} && \multicolumn{2}{c}{Monthly}\\
		&& \multicolumn{1}{c}{$\sigma_m^2$} & \multicolumn{1}{c}{$TO_m$}&& \multicolumn{1}{c}{$\sigma_m^2$} & \multicolumn{1}{c}{$TO_m$}&& \multicolumn{1}{c}{$\sigma_m^2$} & \multicolumn{1}{c}{$TO_m$}\\
		\cline{3-4}
		\cline{6-7}
		\cline{9-10}
		\endhead
		\multicolumn{10}{r}{\textit{Continued on next page}} \\
		\endfoot
		\endlastfoot
		\multirow{11}{*}{\rotatebox{90}{Bull-1}}
		&$1/N$&0.009883&0.005169&&0.009873&0.002335&&0.009907&0.001103\\
		&$NA$&0.000543&0.150112&&0.000794&0.060372&&0.000834&0.026885\\
		&$NA^F$&0.000731&\underline{0.001378}&&0.000732&\textbf{0.000607}&&0.000736&0.000327\\
		&$EWMA$&0.000735&0.088541&&0.000748&0.043884&&0.000754&0.015613\\
		&$PCA$&0.000850&0.236835&&0.000827&0.090803&&0.000795&0.020906\\
		&$RMT$&0.000856&0.150601&&0.000860&0.067402&&0.000826&0.025859\\
		&$LW$&0.001054&0.091979&&0.001572&0.044883&&0.001692&0.022283\\
		&$LW^F$&0.000780&\textbf{0.001228}&&0.000781&\underline{0.000610}&&0.000783&\underline{0.000349}\\
		&$CCC$&0.000620&0.119486&&\underline{0.000681}&0.053221&&\underline{0.000728}&0.013605\\
		&$DCC$&\underline{0.000529}&0.207262&&0.000698&0.093558&&0.000781&0.024324\\
		&$DCC^{NL}$&\textbf{0.000522}&0.217432&&0.000784&0.097809&&0.000875&0.034673\\
		
		&$CAB$&0.000634&0.086489&&\textbf{0.000656}&0.055931&&\textbf{0.000707}&0.023736\\
		\\
		\multirow{11}{*}{\rotatebox{90}{Bear}}
		&$1/N$&0.029939&0.007221&&0.029940&0.003443&&0.029793&0.001599\\
		&$NA$&\textbf{0.002207}&0.092111&&0.002682&0.032018&&0.002959&0.018142\\
		&$NA^F$&0.002891&\textbf{0.001650}&&0.002892&\textbf{0.000935}&&0.002906&\textbf{0.000436}\\
		&$EWMA$&0.002908&0.079065&&0.002843&0.036314&&0.003120&0.019997\\
		&$PCA$&0.002800&0.114256&&0.002792&0.043637&&0.002864&0.011111\\
		&$RMT$&0.002983&0.094396&&0.002896&0.037508&&0.003153&0.023308\\
		&$LW$&0.003118&0.095723&&0.003970&0.041663&&0.004754&0.021610\\
		&$LW^F$&0.003196&\underline{0.002035}&&0.003197&\underline{0.001043}&&0.003201&\underline{0.000570}\\
		&$CCC$&0.002497&0.129487&&\underline{0.002646}&0.043843&&\textbf{0.002590}&0.012451\\
		&$DCC$&0.002355&0.067768&&\textbf{0.002639}&0.015372&&\underline{0.002636}&0.005911\\
		&$DCC^{NL}$&\underline{0.002269}&0.083567&&0.002652&0.024277&&0.002730&0.010308\\
		
		&$CAB$&0.002273&0.030506&&0.002464&0.022686&&0.002656&0.016836\\
		
		\multirow{11}{*}{\rotatebox{90}{Bull-2}}
		&$1/N$&0.019967&0.005666&&0.019956&0.002382&&0.019933&0.001334\\
		&$NA$&\textbf{0.002823}&0.118997&&0.003526&0.053457&&0.003667&0.029686\\
		&$NA^F$&0.003398&\textbf{0.001426}&&0.003400&\textbf{0.000577}&&0.003406&\textbf{0.000261}\\
		&$EWMA$&0.003432&0.073812&&0.003433&0.034460&&0.003630&0.016887\\
		&$PCA$&0.003793&0.194898&&0.003678&0.063899&&0.003597&0.024379\\
		&$RMT$&0.003668&0.118539&&0.003720&0.055498&&0.003762&0.029837\\
		&$LW$&0.003512&0.102531&&0.004274&0.048313&&0.004245&0.020765\\
		&$LW^F$&0.003874&\underline{0.002113}&&0.003879&\underline{0.001123}&&0.003889&\underline{0.000788}\\
		&$CCC$&0.003233&0.129874&&0.003635&0.047812&&0.003671&0.015276\\
		&$DCC$&0.002947&0.071444&&\textbf{0.003294}&0.026871&&\textbf{0.003364}&0.009476\\
		&$DCC^{NL}$&\underline{0.002924}&0.085242&&\underline{0.003318}&0.034070&&\underline{0.003365}&0.012661\\
		
		&$CAB$&0.003323&0.087821&&0.003547&0.052596&&0.003885&0.029487\\
	\end{longtable}
	\endgroup	
	
	Examining the same metrics across distinct market regimes (Table~\ref{tab:portfoliovariancesmr}) reveals that $CAB$ remains a robust contender under varying conditions: in Bull-1, it consistently competes for the lowest variance, even attaining the best performance in weekly and monthly rebalancing; in the Bear regime, $CAB$ ranks near the top in daily and weekly horizons, balancing lower turnover against marginally higher volatility than a few dynamic methods; and while Bull-2 sees other estimators occasionally surpass $CAB$, it still demonstrates stable risk control with moderate trading activity, indicating that its effectiveness persists across both tranquil and turbulent market environments.
	
	Comparisons to the $1/N$ benchmark further highlight the importance of adaptivity. While $1/N$ offers minimal turnover, it endures noticeably higher variance in all market conditions, reflecting its inability to exploit up-to-date correlation information. Indeed, the difference in annualised variances between $CAB$ and $1/N$ can reach 80-90 basis points in certain scenarios, a gap that can substantially affect risk-adjusted returns and capital preservation.
	
	Overall, $CAB$ tends to exhibit moderate turnover, balancing the need to adjust to changing correlations without incurring unnecessary trades. This stability benefits investors who must limit transaction costs or adhere to regulatory constraints on turnover. These results underscore that advanced covariance forecast can yield \emph{economically meaningful} improvements in risk control, particularly over the medium-term horizons favoured by many institutional investors. The $CAB$ model stands out by combining strong variance reduction with stable rebalancing behaviour, balancing responsiveness to market changes and turnover efficiency.
	
	\section{Conclusion}
	\label{sec:conclusion}
	
	We propose a novel DL framework for medium-term covariance forecasting in multi-asset portfolios, combining 3D convolutions, bidirectional LSTMs, and multi-head attention. Empirical tests on various ETFs from 2017 to 2023 reveal that this model, enhanced by a final shrinkage step, consistently outperforms classical benchmarks across different forecast horizons and market regimes.
	
	Notably, its performance advantage remains robust whether one subtracts a risk-free rate from returns, suggesting broad applicability in diverse portfolio management practices.
	
	In portfolio experiments, the proposed method enables GMV strategies to achieve lower out-of-sample volatility with moderate turnover, underlining the tangible economic value of improved covariance forecast. By bridging cutting-edge DL techniques with established financial principles, our work highlights the promise of sophisticated spatiotemporal modelling for risk management and allocation decisions, especially at horizons where structural shifts and evolving correlations pose unique forecasting challenges.
	
	Still, two research avenues warrant further exploration. First, extending the network architecture to accommodate skewed or heavy-tailed distributions could better capture tail risk in times of market stress. Second, explicitly incorporating transaction costs or liquidity constraints within the optimisation process might yield more realistic and implementable trading strategies. 
	
	We hope these findings encourage the finance community to explore and adapt advanced neural architectures for medium-term and long-horizon applications, strengthening the link between machine learning innovation and effective real-world asset allocation.
	
	\clearpage
	
	\bibliographystyle{apalike}
	\bibliography{references}
	
	\clearpage
	\appendix
	\section{Appendix}\label{app:myappendix}
	\renewcommand{\thetable}{\thesection\arabic{table}}
	\renewcommand{\theequation}{\thesection.\arabic{equation}}
	\setcounter{equation}{0}
	\setcounter{table}{0} 
	\subsection{LSTM cell detailed explanation}
	\label{LSTM}
	An LSTM \citep{Hochreiter1997} cell consists of three main gates: the forget gate ($\textbf{\textit{f}}_t$), the input gate ($\textbf{\textit{i}}_t$), and the output gate ($\textbf{\textit{o}}_t$). They are respectively calculated as:
	\begin{equation}
		\textbf{\textit{f}}_t=\sigma(\mathbf{W}_f.[\mathbf{h}_{(t-1)},\mathbf{x}_t\ ]+\mathbf{b}_f )\label{2.32}
	\end{equation}
	\begin{equation}
		\textbf{\textit{i}}_t=\sigma(\mathbf{W}_i.[\mathbf{h}_{(t-1)},\mathbf{x}_t\ ]+\mathbf{b}_i )\label{2.33}
	\end{equation}
	\begin{equation}
		\textbf{\textit{o}}_t=\sigma(\mathbf{W}_o.[\mathbf{h}_{(t-1)},\mathbf{x}_t\ 	]+\mathbf{b}_o )\label{2.34}
	\end{equation}
	where $\mathbf{x}_t$ is the vector input at time $t$, $\mathbf{h}_{(t-1)}$ is the hidden vector state, $\mathbf{W}_f$, $\mathbf{W}_i$ and $\mathbf{W}_c$ are weight matrices, and $\mathbf{b}_f$, $\mathbf{b}_i$ and $\mathbf{b}_o$ are bias vectors. The sigmoid function $\sigma$ ensures that the output values are between 0 and 1, representing how much each component should be forgotten.
	
	The forget gate decides what information should be discarded from the cell state. Next, the input gate determines what new information should be added to the cell state. This process involves two steps: calculating the input gate Eq.~\eqref{2.33} and creating new candidate values:
	\begin{equation}
		\mathbf{\widetilde{C}}_t = \tanh\left( \mathbf{W}_c \cdot \left[ \mathbf{h}_{t-1}, \mathbf{x}_t \right] + \mathbf{b}_c \right) \label{2.35}
	\end{equation}
	where $\mathbf{W}_c$ is the weight matrix, and $\mathbf{b}_c$ is a bias vector. The input gate modulates the extent to which new information is added to the cell state. The cell state update combines the forget gate and input gate operations to update the cell state:
	\begin{equation}
		\mathbf{C}_t=\textbf{\textit{f}}_t\odot\mathbf{C}_(t-1)+\textbf{\textit{i}}_t\odot\mathbf{\widetilde{C}}_t\label{2.36}
	\end{equation}
	
	This equation ensures that the cell state retains essential information over long periods.
	
	Finally, the output gate controls what information should be outputted from the cell. This involves two steps: calculating the output gate Eq.~\eqref{2.34} and determining the hidden state:
	\begin{equation}
		\mathbf{h}_t=\textbf{\textit{o}}_t\odot\tanh{\left(\mathbf{C}_t\right)}\label{2.37}
	\end{equation}
	
	\subsection{Hyperparameter Grid Search}
	\label{hyper}
	\begin{table}[h]
		\centering
		\caption{Hyperparameter Grid Search table.}
		\begin{threeparttable}
			\begin{tabular}{lc}
				Hyperparameter & Range\\
				\hline
				Optimiser & Adam, SGD\\
				Batch size & 32, 64, 128, 256\\
				Learning rate & $10^{-3}$, $10^{-4}$, $10^{-5}$ \\
				Lookback window ($L$) & 20, 40, 60, 80, 100, 120, 250 \\
				3D convolution kernel size ($ks$) & 3, 5, 7\\
				BiLSTM hidden dimensions ($hd$) & 32, 64, 128, 256\\
				BiLSTM stacked layers ($u$) & 3, 4, 5, 6, 7\\
				Number of heads ($he$)\tnote{a} & 2, 4, 6, 8, 16, 32\\
				Shrinkage factor ($\phi$) & 0.00, 0.20, 0.40, 0.60, 0.80, 1.00
			\end{tabular}
		\end{threeparttable}
	\end{table}
	
	\subsection{Extended ETF List}
	\begingroup
		\begin{longtable}{lccc}
			\caption{List of ETFs used in the larger dataset.}\label{tab:listofetfslarge}\\
			Name&Ticker&Incept. Date&Asset Class\\
			\endfirsthead 
			Name&Ticker&Incept. Date&Asset Class\\
			\hline
			\endhead
			\multicolumn{4}{r}{\textit{Continued on next page}} \\
			\endfoot
			\endlastfoot
			\hline
			iShares Core International Aggregate Bond ETF&IAGG&10/11/2015&Fixed Income\\
			iShares MSCI Saudi Arabia ETF&KSA&16/09/2015&Equity\\
			iShares Interest Rate Hedged Long-Term Corporate Bond ETF&IGBH&22/07/2015&Fixed Income\\
			iShares Currency Hedged MSCI ACWI ex U.S. ETF&HAWX&29/06/2015&Equity\\
			iShares Currency Hedged MSCI EAFE Small-Cap ETF&HSCZ&29/06/2015&Equity\\
			iShares MSCI Intl Value Factor ETF&IVLU&16/06/2015&Equity\\
			iShares Convertible Bond ETF&ICVT&02/06/2015&Fixed Income\\
			iShares U.S. Equity Factor ETF&LRGF&28/04/2015&Equity\\
			iShares U.S. Small-Cap Equity Factor ETF&SMLF&28/04/2015&Equity\\
			iShares International Equity Factor ETF&INTF&28/04/2015&Equity\\
			iShares International Small-Cap Equity Factor ETF&ISCF&28/04/2015&Equity\\
			iShares Global Equity Factor ETF&GLOF&28/04/2015&Equity\\
			iShares Exponential Technologies ETF&XT&19/03/2015&Equity\\
			iShares iBonds Dec 2025 Term Corporate ETF&IBDQ&11/03/2015&Fixed Income\\
			iShares Short Maturity Municipal Bond Active ETF&MEAR&03/03/2015&Fixed Income\\
			iShares U.S. Fixed Income Balanced Risk Systematic ETF&FIBR&24/02/2015&Fixed Income\\
			iShares MSCI Intl Quality Factor ETF&IQLT&13/01/2015&Equity\\
			iShares MSCI Intl Momentum Factor ETF&IMTM&13/01/2015&Equity\\
			iShares MSCI ACWI Low Carbon Target ETF&CRBN&08/12/2014&Equity\\
			iShares GSCI Commodity Dynamic Roll Strategy ETF&COMT&15/10/2014&Commodity\\
			iShares Currency Hedged MSCI Emerging Markets ETF&HEEM&23/09/2014&Equity\\
			iShares Currency Hedged MSCI Eurozone ETF&HEZU&09/07/2014&Equity\\
			iShares Global REIT ETF&REET&08/07/2014&Real Estate\\
			iShares Core Total USD Bond Market ETF&IUSB&10/06/2014&Fixed Income\\
			iShares Core Dividend Growth ETF&DGRO&10/06/2014&Equity\\
			iShares Core MSCI Europe ETF&IEUR&10/06/2014&Equity\\
			iShares Core MSCI Pacific ETF&IPAC&10/06/2014&Equity\\
			iShares Interest Rate Hedged Corporate Bond ETF&LQDH&27/05/2014&Fixed Income\\
			iShares Interest Rate Hedged High Yield Bond ETF&HYGH&27/05/2014&Fixed Income\\
			iShares MSCI UAE ETF&UAE&29/04/2014&Equity\\
			iShares MSCI Qatar ETF&QAT&29/04/2014&Equity\\
			iShares Yield Optimized Bond ETF&BYLD&22/04/2014&Fixed Income\\
			iShares Treasury Floating Rate Bond ETF&TFLO&03/02/2014&Fixed Income\\
			iShares Currency Hedged MSCI EAFE ETF&HEFA&31/01/2014&Equity\\
			iShares Currency Hedged MSCI Japan ETF&HEWJ&31/01/2014&Equity\\
			iShares Ultra Short-Term Bond Active ETF&ICSH&11/12/2013&Fixed Income\\
			iShares 0-5 Year High Yield Corporate Bond ETF&SHYG&15/10/2013&Fixed Income\\
			iShares 0-5 Year Investment Grade Corporate Bond ETF&SLQD&15/10/2013&Fixed Income\\
			iShares Short Duration Bond Active ETF&NEAR&25/09/2013&Fixed Income\\
			iShares MSCI USA Quality Factor ETF&QUAL&16/07/2013&Equity\\
			iShares MSCI USA Momentum Factor ETF&MTUM&16/04/2013&Equity\\
			iShares MSCI USA Value Factor ETF&VLUE&16/04/2013&Equity\\
			iShares MSCI USA Size Factor ETF&SIZE&16/04/2013&Equity\\
			iShares Core MSCI EAFE ETF&IEFA&18/10/2012&Equity\\
			iShares Core MSCI Emerging Markets ETF&IEMG&18/10/2012&Equity\\
			iShares Core MSCI Total International Stock ETF&IXUS&18/10/2012&Equity\\
			iShares Core 1-5 Year USD Bond ETF&ISTB&18/10/2012&Fixed Income\\
			iShares J.P. Morgan EM Corporate Bond ETF&CEMB&17/04/2012&Fixed Income\\
			iShares J.P. Morgan EM High Yield Bond ETF&EMHY&03/04/2012&Fixed Income\\
			iShares US \& Intl High Yield Corp Bond ETF&GHYG&03/04/2012&Fixed Income\\
			iShares Morningstar Multi-Asset Income ETF&IYLD&03/04/2012&Multi Asset\\
			iShares International High Yield Bond ETF&HYXU&03/04/2012&Fixed Income\\
			iShares Emerging Markets Dividend ETF&DVYE&23/02/2012&Equity\\
			iShares Asia/Pacific Dividend ETF&DVYA&23/02/2012&Equity\\
			iShares U.S. Treasury Bond ETF&GOVT&14/02/2012&Fixed Income\\
			iShares Aaa - A Rated Corporate Bond ETF&QLTA&14/02/2012&Fixed Income\\
			iShares CMBS ETF&CMBS&14/02/2012&Others\\
			iShares GNMA Bond ETF&GNMA&14/02/2012&Fixed Income\\
			iShares MSCI India Small-Cap ETF&SMIN&08/02/2012&Equity\\
			iShares MSCI Emerging Markets Asia ETF&EEMA&08/02/2012&Equity\\
			iShares MSCI India ETF&INDA&02/02/2012&Equity\\
			iShares MSCI Global Gold Miners ETF&RING&31/01/2012&Equity\\
			iShares MSCI Global Metals \& Mining Producers ETF&PICK&31/01/2012&Equity\\
			iShares MSCI Global Silver and Metals Miners ETF&SLVP&31/01/2012&Equity\\
			iShares MSCI Agriculture Producers ETF&VEGI&31/01/2012&Equity\\
			iShares MSCI Global Energy Producers ETF&FILL&31/01/2012&Equity\\
			iShares MSCI Denmark ETF&EDEN&25/01/2012&Equity\\
			iShares MSCI United Kingdom Small-Cap ETF&EWUS&25/01/2012&Equity\\
			iShares MSCI Finland ETF&EFNL&25/01/2012&Equity\\
			iShares MSCI Norway ETF&ENOR&23/01/2012&Equity\\
			iShares MSCI World ETF&URTH&10/01/2012&Equity\\
			iShares MSCI USA Min Vol Factor ETF&USMV&18/10/2011&Equity\\
			iShares MSCI EAFE Min Vol Factor ETF&EFAV&18/10/2011&Equity\\
			iShares MSCI Global Min Vol Factor ETF&ACWV&18/10/2011&Equity\\
			iShares MSCI Emerging Markets Min Vol Factor ETF&EEMV&18/10/2011&Equity\\
			iShares J.P. Morgan EM Local Currency Bond ETF&LEMB&18/10/2011&Fixed Income\\
			iShares MSCI Emerging Markets Small-Cap ETF&EEMS&16/08/2011&Equity\\
			iShares Floating Rate Bond ETF&FLOT&14/06/2011&Fixed Income\\
			iShares Core High Dividend ETF&HDV&29/03/2011&Equity\\
			iShares MSCI China ETF&MCHI&29/03/2011&Equity\\
			iShares 0-5 Year TIPS Bond ETF&STIP&01/12/2010&Fixed Income\\
			iShares MSCI Brazil Small-Cap ETF&EWZS&28/09/2010&Equity\\
			iShares MSCI Philippines ETF&EPHE&28/09/2010&Equity\\
			iShares MSCI China Small-Cap ETF&ECNS&28/09/2010&Equity\\
			iShares MSCI New Zealand ETF&ENZL&01/09/2010&Equity\\
			iShares MSCI Poland ETF&EPOL&25/05/2010&Equity\\
			iShares MSCI USA Equal Weighted ETF&EUSA&05/05/2010&Equity\\
			iShares MSCI Indonesia ETF&EIDO&05/05/2010&Equity\\
			iShares MSCI Ireland ETF&EIRL&05/05/2010&Equity\\
			iShares MSCI Europe Financials ETF&EUFN&20/01/2010&Equity\\
			iShares 10+ Year Investment Grade Corporate Bond ETF&IGLB&08/12/2009&Fixed Income\\
			iShares Core 10+ Year USD Bond ETF&ILTB&08/12/2009&Fixed Income\\
			iShares India 50 ETF&INDY&18/11/2009&Equity\\
			iShares Russell Top 200 Growth ETF&IWY&22/09/2009&Equity\\
			iShares Russell Top 200 Value ETF&IWX&22/09/2009&Equity\\
			iShares Russell Top 200 ETF&IWL&22/09/2009&Equity\\
			iShares MSCI Peru and Global Exposure ETF&EPU&19/06/2009&Equity\\
			iShares Emerging Markets Infrastructure ETF&EMIF&16/06/2009&Equity\\
			iShares International Treasury Bond ETF&IGOV&21/01/2009&Fixed Income\\
			iShares 1-3 Year International Treasury Bond ETF&ISHG&21/01/2009&Fixed Income\\
			iShares Short-Term National Muni Bond ETF&SUB&05/11/2008&Fixed Income\\
			iShares Agency Bond ETF&AGZ&05/11/2008&Fixed Income\\
			iShares Core Growth Allocation ETF&AOR&04/11/2008&Multi Asset\\
			iShares Core Aggressive Allocation ETF&AOA&04/11/2008&Multi Asset\\
			iShares Core Moderate Allocation ETF&AOM&04/11/2008&Multi Asset\\
			iShares Core Conservative Allocation ETF&AOK&04/11/2008&Multi Asset\\
			iShares MSCI All Country Asia ex Japan ETF&AAXJ&13/08/2008&Equity\\
			iShares Global Clean Energy ETF&ICLN&24/06/2008&Equity\\
			iShares Global Timber \& Forestry ETF&WOOD&24/06/2008&Equity\\
			iShares MSCI ACWI ETF&ACWI&26/03/2008&Equity\\
			iShares MSCI ACWI ex U.S. ETF&ACWX&26/03/2008&Equity\\
			iShares MSCI Israel ETF&EIS&26/03/2008&Equity\\
			iShares MSCI Thailand ETF&THD&26/03/2008&Equity\\
			iShares MSCI Turkey ETF&TUR&26/03/2008&Equity\\
			iShares MSCI Japan Small-Cap ETF&SCJ&20/12/2007&Equity\\
			iShares J.P. Morgan USD Emerging Markets Bond ETF&EMB&17/12/2007&Fixed Income\\
			iShares MSCI EAFE Small-Cap ETF&SCZ&10/12/2007&Equity\\
			iShares Global Infrastructure ETF&IGF&10/12/2007&Equity\\
			iShares MSCI Kokusai ETF&TOK&10/12/2007&Equity\\
			iShares Asia 50 ETF&AIA&13/11/2007&Equity\\
			iShares MSCI Chile ETF&ECH&12/11/2007&Equity\\
			iShares International Developed Real Estate ETF&IFGL&12/11/2007&Real Estate\\
			iShares MSCI Europe Small-Cap ETF&IEUS&12/11/2007&Equity\\
			iShares MSCI BIC ETF&BKF&12/11/2007&Equity\\
			iShares California Muni Bond ETF&CMF&04/10/2007&Fixed Income\\
			iShares New York Muni Bond ETF&NYF&04/10/2007&Fixed Income\\
			iShares National Muni Bond ETF&MUB&07/09/2007&Fixed Income\\
			iShares International Select Dividend ETF&IDV&11/06/2007&Equity\\
			iShares Core U.S. REIT ETF&USRT&01/05/2007&Real Estate\\
			iShares Residential and Multisector Real Estate ETF&REZ&01/05/2007&Real Estate\\
			iShares Mortgage Real Estate ETF&REM&01/05/2007&Real Estate\\
			iShares iBoxx High Yield Corporate Bond ETF&HYG&04/04/2007&Fixed Income\\
			iShares Preferred and Income Securities ETF&PFF&26/03/2007&Equity\\
			iShares MBS ETF&MBB&13/03/2007&Fixed Income\\
			iShares 1-5 Year Investment Grade Corporate Bond ETF&IGSB&05/01/2007&Fixed Income\\
			iShares Short Treasury Bond ETF&SHV&05/01/2007&Fixed Income\\
			iShares 3-7 Year Treasury Bond ETF&IEI&05/01/2007&Fixed Income\\
			iShares 5-10 Year Investment Grade Corporate Bond ETF&IGIB&05/01/2007&Fixed Income\\
			iShares Broad USD Investment Grade Corporate Bond ETF&USIG&05/01/2007&Fixed Income\\
			iShares 10-20 Year Treasury Bond ETF&TLH&05/01/2007&Fixed Income\\
			iShares Intermediate Government/Credit Bond ETF&GVI&05/01/2007&Fixed Income\\
			iShares Government/Credit Bond ETF&GBF&05/01/2007&Fixed Income\\
			iShares MSCI KLD 400 Social ETF&DSI&14/11/2006&Equity\\
			iShares Global Industrials ETF&EXI&12/09/2006&Equity\\
			iShares Global Consumer Staples ETF&KXI&12/09/2006&Equity\\
			iShares Global Consumer Discretionary ETF&RXI&12/09/2006&Equity\\
			iShares Global Materials ETF&MXI&12/09/2006&Equity\\
			iShares Global Utilities ETF&JXI&12/09/2006&Equity\\
			iShares S\&P GSCI Commodity-Indexed Trust&GSG&10/07/2006&Commodity\\
			iShares U.S. Aerospace \& Defense ETF&ITA&01/05/2006&Equity\\
			iShares U.S. Medical Devices ETF&IHI&01/05/2006&Equity\\
			iShares U.S. Home Construction ETF&ITB&01/05/2006&Equity\\
			iShares U.S. Broker-Dealers \& Securities Exchanges ETF&IAI&01/05/2006&Equity\\
			iShares U.S. Regional Banks ETF&IAT&01/05/2006&Equity\\
			iShares U.S. Insurance ETF&IAK&01/05/2006&Equity\\
			iShares U.S. Healthcare Providers ETF&IHF&01/05/2006&Equity\\
			iShares U.S. Pharmaceuticals ETF&IHE&01/05/2006&Equity\\
			iShares U.S. Oil \& Gas Exploration \& Production ETF&IEO&01/05/2006&Equity\\
			iShares U.S. Oil Equipment \& Services ETF&IEZ&01/05/2006&Equity\\
			iShares Silver Trust&SLV&21/04/2006&Commodity\\
			iShares Micro-Cap ETF&IWC&12/08/2005&Equity\\
			iShares MSCI EAFE Value ETF&EFV&01/08/2005&Equity\\
			iShares MSCI EAFE Growth ETF&EFG&01/08/2005&Equity\\
			iShares MSCI USA ESG Select ETF&SUSA&24/01/2005&Equity\\
			iShares Gold Trust&IAU&21/01/2005&Commodity\\
			iShares China Large-Cap ETF&FXI&05/10/2004&Equity\\
			iShares Morningstar Mid-Cap Growth ETF&IMCG&28/06/2004&Equity\\
			iShares Morningstar Growth ETF&ILCG&28/06/2004&Equity\\
			iShares Morningstar U.S. Equity ETF&ILCB&28/06/2004&Equity\\
			iShares Morningstar Mid-Cap ETF&IMCB&28/06/2004&Equity\\
			iShares Morningstar Value ETF&ILCV&28/06/2004&Equity\\
			iShares Morningstar Mid-Cap Value ETF&IMCV&28/06/2004&Equity\\
			iShares Morningstar Small-Cap Growth ETF&ISCG&28/06/2004&Equity\\
			iShares Morningstar Small-Cap Value ETF&ISCV&28/06/2004&Equity\\
			iShares Morningstar Small-Cap ETF&ISCB&28/06/2004&Equity\\
			iShares Core S\&P Total U.S. Stock Market ETF&ITOT&20/01/2004&Equity\\
			iShares TIPS Bond ETF&TIP&04/12/2003&Fixed Income\\
			iShares Select Dividend ETF&DVY&03/11/2003&Equity\\
			iShares U.S. Transportation ETF&IYT&06/10/2003&Equity\\
			iShares Core U.S. Aggregate Bond ETF&AGG&22/09/2003&Fixed Income\\
			iShares MSCI Emerging Markets ETF&EEM&07/04/2003&Equity\\
			iShares MSCI South Africa ETF&EZA&03/02/2003&Equity\\
			iShares 20+ Year Treasury Bond ETF&TLT&22/07/2002&Fixed Income\\
			iShares 7-10 Year Treasury Bond ETF&IEF&22/07/2002&Fixed Income\\
			iShares iBoxx Investment Grade Corporate Bond ETF&LQD&22/07/2002&Fixed Income\\
			iShares 1-3 Year Treasury Bond ETF&SHY&22/07/2002&Fixed Income\\
			iShares Global Healthcare ETF&IXJ&13/11/2001&Equity\\
			iShares Global Tech ETF&IXN&12/11/2001&Equity\\
			iShares Global Energy ETF&IXC&12/11/2001&Equity\\
			iShares Global Financials ETF&IXG&12/11/2001&Equity\\
			iShares Global Comm Services ETF&IXP&12/11/2001&Equity\\
			iShares MSCI Pacific ex Japan ETF&EPP&25/10/2001&Equity\\
			iShares Latin America 40 ETF&ILF&25/10/2001&Equity\\
			iShares JPX-Nikkei 400 ETF&JPXN&23/10/2001&Equity\\
			iShares North American Natural Resources ETF&IGE&22/10/2001&Equity\\
			iShares MSCI EAFE ETF&EFA&14/08/2001&Equity\\
			iShares Russell Mid-Cap ETF&IWR&17/07/2001&Equity\\
			iShares Russell Mid-Cap Growth ETF&IWP&17/07/2001&Equity\\
			iShares Russell Mid-Cap Value ETF&IWS&17/07/2001&Equity\\
			iShares Semiconductor ETF&SOXX&10/07/2001&Equity\\
			iShares Expanded Tech-Software Sector ETF&IGV&10/07/2001&Equity\\
			iShares U.S. Digital Infrastructure and Real Estate ETF&IDGT&10/07/2001&Equity\\
			iShares Expanded Tech Sector ETF&IGM&13/03/2001&Equity\\
			iShares Biotechnology ETF&IBB&05/02/2001&Equity\\
			iShares Cohen \& Steers REIT ETF&ICF&29/01/2001&Real Estate\\
			iShares Global 100 ETF&IOO&05/12/2000&Equity\\
			iShares S\&P 100 ETF&OEF&23/10/2000&Equity\\
			iShares MSCI Eurozone ETF&EZU&25/07/2000&Equity\\
			iShares Europe ETF&IEV&25/07/2000&Equity\\
			iShares Core S\&P U.S. Growth ETF&IUSG&24/07/2000&Equity\\
			iShares Core S\&P U.S. Value ETF&IUSV&24/07/2000&Equity\\
			iShares Russell 2000 Growth ETF&IWO&24/07/2000&Equity\\
			iShares Russell 2000 Value ETF&IWN&24/07/2000&Equity\\
			iShares S\&P Mid-Cap 400 Growth ETF&IJK&24/07/2000&Equity\\
			iShares S\&P Mid-Cap 400 Value ETF&IJJ&24/07/2000&Equity\\
			iShares S\&P Small-Cap 600 Value ETF&IJS&24/07/2000&Equity\\
			iShares S\&P Small-Cap 600 Growth ETF&IJT&24/07/2000&Equity\\
			iShares MSCI Brazil ETF&EWZ&10/07/2000&Equity\\
			iShares MSCI Taiwan ETF&EWT&20/06/2000&Equity\\
			iShares U.S. Real Estate ETF&IYR&12/06/2000&Real Estate\\
			iShares U.S. Healthcare ETF&IYH&12/06/2000&Equity\\
			iShares Dow Jones U.S. ETF&IYY&12/06/2000&Equity\\
			iShares U.S. Consumer Discretionary ETF&IYC&12/06/2000&Equity\\
			iShares U.S. Financial Services ETF&IYG&12/06/2000&Equity\\
			iShares U.S. Industrials ETF&IYJ&12/06/2000&Equity\\
			iShares U.S. Energy ETF&IYE&12/06/2000&Equity\\
			iShares U.S. Consumer Staples ETF&IYK&12/06/2000&Equity\\
			iShares U.S. Utilities ETF&IDU&12/06/2000&Equity\\
			iShares U.S. Basic Materials ETF&IYM&12/06/2000&Equity\\
			iShares Russell 1000 Growth ETF&IWF&22/05/2000&Equity\\
			iShares Core S\&P Mid-Cap ETF&IJH&22/05/2000&Equity\\
			iShares Core S\&P Small-Cap ETF&IJR&22/05/2000&Equity\\
			iShares Russell 2000 ETF&IWM&22/05/2000&Equity\\
			iShares Russell 1000 Value ETF&IWD&22/05/2000&Equity\\
			iShares S\&P 500 Growth ETF&IVW&22/05/2000&Equity\\
			iShares S\&P 500 Value ETF&IVE&22/05/2000&Equity\\
			iShares Russell 3000 ETF&IWV&22/05/2000&Equity\\
			iShares U.S. Financials ETF&IYF&22/05/2000&Equity\\
			iShares U.S. Telecommunications ETF&IYZ&22/05/2000&Equity\\
			iShares Core S\&P 500 ETF&IVV&15/05/2000&Equity\\
			iShares Russell 1000 ETF&IWB&15/05/2000&Equity\\
			iShares U.S. Technology ETF&IYW&15/05/2000&Equity\\
			iShares MSCI South Korea ETF&EWY&09/05/2000&Equity\\
			iShares MSCI Japan ETF&EWJ&12/03/1996&Equity\\
			iShares MSCI United Kingdom ETF&EWU&12/03/1996&Equity\\
			iShares MSCI Canada ETF&EWC&12/03/1996&Equity\\
			iShares MSCI Australia ETF&EWA&12/03/1996&Equity\\
			iShares MSCI Mexico ETF&EWW&12/03/1996&Equity\\
			iShares MSCI Switzerland ETF&EWL&12/03/1996&Equity\\
			iShares MSCI Germany ETF&EWG&12/03/1996&Equity\\
			iShares MSCI Spain ETF&EWP&12/03/1996&Equity\\
			iShares MSCI Singapore ETF&EWS&12/03/1996&Equity\\
			iShares MSCI Hong Kong ETF&EWH&12/03/1996&Equity\\
			iShares MSCI France ETF&EWQ&12/03/1996&Equity\\
			iShares MSCI Italy ETF&EWI&12/03/1996&Equity\\
			iShares MSCI Sweden ETF&EWD&12/03/1996&Equity\\
			iShares MSCI Malaysia ETF&EWM&12/03/1996&Equity\\
			iShares MSCI Netherlands ETF&EWN&12/03/1996&Equity\\
			iShares MSCI Austria ETF&EWO&12/03/1996&Equity\\
			iShares MSCI Belgium ETF&EWK&12/03/1996&Equity
		\end{longtable}
		\subsection{Extended Tables}
		\begingroup
		\begin{longtable}{llrrrrrrrrrrr}
			\caption{Average Euclidean ($\mathcal{L}^E$) and Frobenius ($\mathcal{L}^F$) distances ($\times 10^5$) using \emph{excess} returns with different horizons ($F$) for different testing periods: Overall (1~January~2021 to 31~December~2023), Bull-1 (1~January~2021 to 2~January~2022), Bear (3~January~2022 to 12~June~2022), and Bull-2 (13~June~2022 to 31~December~2023).}\label{tab:diffhorizonsexc}\\
			&&\multicolumn{2}{c}{Overall}&&\multicolumn{2}{c}{Bull-1}&&\multicolumn{2}{c}{Bear}&&\multicolumn{2}{c}{Bull-2}\\
			&&\multicolumn{1}{c}{$\mathcal{L}^E$}&\multicolumn{1}{c}{$\mathcal{L}^F$}&&\multicolumn{1}{c}{$\mathcal{L}^E$}&\multicolumn{1}{c}{$\mathcal{L}^F$}&&\multicolumn{1}{c}{$\mathcal{L}^E$}&\multicolumn{1}{c}{$\mathcal{L}^F$}&&\multicolumn{1}{c}{$\mathcal{L}^E$}&\multicolumn{1}{c}{$\mathcal{L}^F$}\\
			\cline{3-4}
			\cline{6-7}
			\cline{9-10}
			\cline{12-13}
			\endfirsthead 
			&&\multicolumn{1}{c}{$\mathcal{L}^E$}&\multicolumn{1}{c}{$\mathcal{L}^F$}&&\multicolumn{1}{c}{$\mathcal{L}^E$}&\multicolumn{1}{c}{$\mathcal{L}^F$}&&\multicolumn{1}{c}{$\mathcal{L}^E$}&\multicolumn{1}{c}{$\mathcal{L}^F$}&&\multicolumn{1}{c}{$\mathcal{L}^E$}&\multicolumn{1}{c}{$\mathcal{L}^F$}\\
			\cline{3-4}
			\cline{6-7}
			\cline{9-10}
			\cline{12-13}
			\endhead
			\multicolumn{13}{r}{\textit{Continued on next page}} \\
			\endfoot
			\endlastfoot
			\multirow{11}{*}{\rotatebox{90}{$F=10$}}
			&$NA$&70.309&107.739&&54.194&82.600&&107.447&165.092&&69.960&107.363\\
			&$NA^F$&62.260&95.586&&46.956&71.921&&106.652&163.471&&59.271&91.185\\
			&$EWMA$&58.851&90.253&&\textbf{42.125}&\textbf{64.612}&&\underline{94.704}&\textbf{145.253}&&59.272&90.889\\
			&$PCA$&71.266&109.164&&54.984&83.752&&110.811&170.301&&70.323&107.864\\
			&$RMT$&71.064&108.951&&54.819&83.588&&110.623&170.011&&70.094&107.641\\
			&$LW$&63.548&97.681&&47.376&72.757&&104.804&160.926&&62.036&95.449\\
			&$LW^F$&75.861&115.370&&69.194&104.027&&105.832&162.356&&71.461&109.054\\
			&$CCC$&\underline{58.111}&\underline{89.824}&&\underline{42.451}&\underline{65.818}&&95.819&147.003&&\underline{57.301}&\underline{88.763}\\
			&$DCC$&62.210&95.018&&48.557&73.797&&96.913&147.989&&60.972&93.374\\
			&$DCC^{NL}$&65.361&99.154&&51.461&77.640&&100.48&152.649&&64.160&97.548\\
			&$CAB$&\textbf{56.954}&\textbf{87.317}&&44.134&67.401&&\textbf{94.689}&\underline{145.287}&&\textbf{54.291}&\textbf{83.369}\\
			\\
			\\
			\\
			\multirow{11}{*}{\rotatebox{90}{$F=40$}}
			&$NA$&45.872&70.272&&24.755&38.850&&\underline{72.135}&\underline{108.838}&&51.939&79.449\\
			&$NA^F$&47.436&72.906&&31.056&47.811&&90.540&139.813&&45.522&69.719\\
			&$EWMA$&44.734&68.528&&26.175&40.546&&74.009&112.405&&48.262&73.923\\
			&$PCA$&46.390&71.007&&24.934&39.238&&72.294&109.567&&52.781&80.410\\
			&$RMT$&46.177&70.771&&24.796&39.028&&72.411&109.269&&52.424&80.176\\
			&$LW$&43.441&66.728&&2\underline{3.834}&\underline{37.275}&&76.162&114.702&&46.645&71.886\\
			&$LW^F$&63.848&96.730&&62.374&93.125&&78.628&122.264&&60.501&91.635\\
			&$CCC$&43.434&67.183&&27.608&43.493&&75.190&114.891&&44.465&68.675\\
			&$DCC$&\underline{42.622}&\underline{66.124}&&27.194&42.996&&73.983&113.348&&\underline{43.509}&\underline{67.391}\\
			&$DCC^{NL}$&43.764&67.604&&28.135&44.201&&75.173&114.919&&44.767&69.024\\
			&$CAB$&\textbf{22.851}&\textbf{35.411}&&\textbf{13.091}&\textbf{20.268}&&\textbf{42.372}&\textbf{65.754}&&\textbf{23.505}&\textbf{36.408}\\
			\\
			\\
			\multirow{11}{*}{\rotatebox{90}{$F=60$}}
			&$NA$&44.053&68.114&&26.115&42.593&&72.110&109.681&&47.531&72.584\\
			&$NA^F$&45.519&69.961&&30.907&47.659&&83.084&128.667&&44.068&67.348\\
			&$EWMA$&44.704&68.490&&26.626&41.728&&74.557&112.995&&47.751&72.910\\
			&$PCA$&44.550&68.801&&26.466&43.142&&72.705&111.010&&48.096&73.173\\
			&$RMT$&44.334&68.575&&26.291&42.936&&72.803&110.768&&47.761&72.940\\
			&$LW$&42.035&65.084&&\underline{24.836}&\underline{40.385}&&75.018&113.948&&43.600&66.895\\
			&$LW^F$&62.001&93.989&&62.649&93.590&&69.286&108.383&&59.458&90.055\\
			&$CCC$&42.647&66.037&&29.089&45.917&&\underline{68.525}&\underline{105.242}&&43.916&67.686\\
			&$DCC$&\underline{40.986}&\underline{63.934}&&26.426&42.585&&69.459&106.517&&\underline{42.150}&\underline{65.398}\\
			&$DCC^{NL}$&42.268&65.595&&27.106&43.443&&71.582&109.283&&43.579&67.259\\
			
			&$CAB$&\textbf{22.305}&\textbf{34.990}&&\textbf{12.842}&\textbf{20.226}&&\textbf{39.166}&\textbf{62.879}&&\textbf{23.540}&\textbf{36.458}\\
			\\
			\multirow{5}{*}{\rotatebox{90}{$F=90$}}
			&$NA$&43.972&68.530&&30.179&49.646&&76.424&117.032&&43.479&66.670\\
			&$NA^F$&43.572&67.004&&30.527&47.207&&83.143&128.419&&40.520&\underline{61.975}\\
			&$EWMA$&44.720&68.781&&28.925&\underline{45.742}&&73.402&111.890&&46.624&71.190\\
			&$PCA$&44.464&69.171&&30.520&50.091&&76.867&118.267&&44.085&67.265\\
			&$RMT$&44.239&68.951&&30.343&49.925&&77.105&118.060&&43.693&67.005\\
			\multirow{6}{*}{\rotatebox{90}{$F=90$}}
			&$LW$&42.446&66.204&&\underline{28.694}&47.283&&79.704&121.665&&40.527&62.340\\
			&$LW^F$&60.006&91.073&&61.704&92.289&&\underline{64.055}&\underline{100.713}&&57.724&87.477\\
			&$CCC$&43.390&67.342&&32.735&51.757&&68.495&105.076&&42.999&66.475\\
			&$DCC$&\underline{40.234}&\underline{63.342}&&29.961&48.303&&65.688&101.473&&\underline{39.492}&62.005\\
			&$DCC^{NL}$&42.615&66.462&&29.762&48.148&&71.380&108.855&&42.584&66.011\\
			
			&$CAB$&\textbf{13.977}&\textbf{21.752}&&\textbf{10.080}&\textbf{16.184}&&\textbf{20.742}&\textbf{32.133}&&\textbf{14.538}&\textbf{22.346}\\
			\\
			\multirow{11}{*}{\rotatebox{90}{$F=120$}}
			&$NA$&48.590&\underline{75.398}&&\underline{36.397}&\underline{59.019}&&84.342&128.425&&\underline{46.097}&\underline{70.593}\\
			&$NA^F$&41.909&64.465&&31.713&49.127&&84.615&129.958&&36.093&55.352\\
			&$EWMA$&45.330&69.861&&33.556&53.195&&71.953&109.497&&45.222&69.142\\
			&$PCA$&49.115&76.015&&36.715&59.374&&84.740&129.651&&46.794&71.202\\
			&$RMT$&48.867&75.827&&36.519&59.220&&85.041&129.486&&46.352&70.985\\
			&$LW$&47.098&73.175&&34.956&56.786&&87.296&132.635&&43.277&66.502\\
			&$LW^F$&58.216&88.420&&60.114&90.286&&\underline{64.569}&\underline{100.736}&&55.133&83.621\\
			&$CCC$&44.163&68.631&&36.440&57.692&&71.518&108.718&&41.212&64.060\\
			&$DCC$&\underline{40.946}&64.563&&33.741&54.358&&68.094&104.261&&37.719&59.629\\
			&$DCC^{NL}$&44.082&68.671&&34.120&54.937&&74.174&112.239&&41.789&64.902\\
			&$CAB$&\textbf{12.635}&\textbf{19.816}&&\textbf{10.718}&\textbf{17.319}&&\textbf{18.659}&\textbf{29.190}&&\textbf{12.125}&\textbf{18.707}\\
			\\
			\multirow{11}{*}{\rotatebox{90}{$F=180$}}
			&$NA$&57.475&88.882&&50.083&79.438&&92.682&140.191&&52.022&80.073\\
			&$NA^F$&\underline{38.460}&\underline{59.180}&&\underline{32.171}&\underline{49.910}&&82.117&124.805&&\underline{29.829}&\underline{46.088}\\
			&$EWMA$&50.035&77.019&&43.437&68.307&&76.903&116.122&&46.487&71.288\\
			&$PCA$&58.026&89.446&&50.494&79.976&&92.884&141.079&&52.765&80.559\\
			&$RMT$&57.744&89.294&&50.351&79.843&&93.189&140.964&&52.223&80.384\\
			&$LW$&56.137&86.937&&49.005&77.757&&94.604&142.932&&49.567&76.591\\
			&$LW^F$&54.308&82.570&&54.261&81.972&&\underline{45.646}&\underline{95.111}&&52.187&79.306\\
			&$CCC$&44.949&70.232&&40.531&64.242&&72.901&110.150&&39.678&62.498\\
			&$DCC$&46.528&72.381&&41.217&65.244&&75.945&114.305&&41.410&64.807\\
			&$DCC^{NL}$&46.532&72.386&&41.220&65.248&&75.950&114.312&&41.414&64.812\\
			
			&$CAB$&\textbf{12.356}&\textbf{19.146}&&\textbf{11.116}&\textbf{17.602}&&\textbf{18.159}&\textbf{27.468}&&\textbf{11.472}&\textbf{17.726}\\
			\\
			\multirow{11}{*}{\rotatebox{90}{$F=250$}}
			&$NA$&76.850&117.44&&107.237&163.541&&79.505&120.16&&56.341&86.706\\
			&$NA^F$&\underline{34.002}&\underline{52.367}&&\underline{33.076}&\underline{51.123}&&64.126&97.309&&25.830&\underline{40.087}\\
			&$PCA$&77.628&118.284&&108.476&165.211&&79.544&120.791&&57.035&87.075\\
			&$RMT$&77.315&118.139&&108.26&165.074&&79.841&120.668&&56.481&86.919\\
			&$EWMA$&53.737&82.720&&53.030&82.554&&72.648&109.762&&48.689&74.953\\
			&$LW$&74.903&114.697&&103.984&158.986&&80.825&122.029&&54.291&83.797\\
			&$LW^F$&49.653&75.499&&47.293&71.542&&\underline{45.646}&\underline{70.522}&&52.353&79.519\\
			&$CCC$&43.487&68.479&&42.628&67.459&&58.966&89.895&&39.537&62.904\\
			&$DCC$&40.299&64.494&&38.892&62.869&&53.375&82.622&&37.405&60.269\\
			&$DCC^{NL}$&40.223&66.604&&49.514&77.717&&81.610&120.697&&\underline{22.135}&43.633\\
			&$CAB$&\textbf{16.116}&\textbf{24.758}&&\textbf{22.511}&\textbf{34.543}&&\textbf{16.201}&\textbf{24.572}&&\textbf{11.938}&\textbf{18.457}\\
		\end{longtable}
		\endgroup
		\begingroup
		\begin{longtable}{llrrrrrrrrrrr}
			\caption{Average Euclidean ($\mathcal{L}^E$) and Frobenius ($\mathcal{L}^F$) distances ($\times 10^5$) using \emph{raw} returns with different horizons ($F$) for different testing periods: Overall (1~January~2021 to 31~December~2023), Bull-1 (1~January~2021 to 2~January~2022), Bear (3~January~2022 to 12~June~2022), and Bull-2 (13~June~2022 to 31~December~2023).}\label{tab:diffhorizonsraw}\\
			&&\multicolumn{2}{c}{Overall}&&\multicolumn{2}{c}{Bull-1}&&\multicolumn{2}{c}{Bear}&&\multicolumn{2}{c}{Bull-2}\\
			&&\multicolumn{1}{c}{$\mathcal{L}^E$}&\multicolumn{1}{c}{$\mathcal{L}^F$}&&\multicolumn{1}{c}{$\mathcal{L}^E$}&\multicolumn{1}{c}{$\mathcal{L}^F$}&&\multicolumn{1}{c}{$\mathcal{L}^E$}&\multicolumn{1}{c}{$\mathcal{L}^F$}&&\multicolumn{1}{c}{$\mathcal{L}^E$}&\multicolumn{1}{c}{$\mathcal{L}^F$}\\
			\cline{3-4}
			\cline{6-7}
			\cline{9-10}
			\cline{12-13}
			\endfirsthead 
			&&\multicolumn{1}{c}{$\mathcal{L}^E$}&\multicolumn{1}{c}{$\mathcal{L}^F$}&&\multicolumn{1}{c}{$\mathcal{L}^E$}&\multicolumn{1}{c}{$\mathcal{L}^F$}&&\multicolumn{1}{c}{$\mathcal{L}^E$}&\multicolumn{1}{c}{$\mathcal{L}^F$}&&\multicolumn{1}{c}{$\mathcal{L}^E$}&\multicolumn{1}{c}{$\mathcal{L}^F$}\\
			\cline{3-4}
			\cline{6-7}
			\cline{9-10}
			\cline{12-13}
			\endhead
			\multicolumn{13}{r}{\textit{Continued on next page}} \\
			\endfoot
			\endlastfoot
			\multirow{11}{*}{\rotatebox{90}{$F=10$}}
			&$NA$&70.309&107.739&&54.193&82.599&&107.449&165.094&&69.959&107.362\\
			&$NA^F$&62.057&95.297&&46.340&71.028&&107.068&164.102&&59.155&91.020\\
			&$EWMA$&58.908&90.338&&\textbf{42.125}&\textbf{64.612}&&\underline{94.704}&\underline{145.253}&&59.384&91.053\\
			&$PCA$&71.265&109.164&&54.983&83.750&&110.813&170.304&&70.322&107.863\\
			&$RMT$&71.064&108.951&&54.818&83.587&&110.625&170.014&&70.093&107.640\\
			&$LW$&63.547&97.680&&47.376&72.755&&104.804&160.927&&62.035&95.448\\
			&$LW^F$&75.854&115.361&&69.181&104.007&&105.831&162.355&&71.457&109.048\\
			&$CCC$&\textbf{58.092}&\underline{89.783}&&\underline{42.384}&\underline{65.733}&&95.843&147.035&&57.300&88.730\\
			&$DCC$&62.186&94.973&&48.470&73.686&&96.894&147.968&&60.986&93.364\\
			&$DCC^{NL}$&65.338&99.111&&51.367&77.516&&100.447&152.608&&64.186&97.557\\
			
			&$CAB$&\underline{58.272}&\textbf{89.475}&&43.967&67.109&&\textbf{94.554}&\textbf{145.032}&&\textbf{56.996}&\textbf{87.822}\\
			\\
			\multirow{11}{*}{\rotatebox{90}{$F=40$}}
			&$NA$&45.876&70.278&&24.754&38.850&&\underline{72.131}&\underline{108.833}&&51.948&79.462\\
			&$NA^F$&47.200&72.577&&30.161&46.526&&91.426&141.121&&45.386&69.534\\
			&$EWMA$&44.796&68.618&&26.174&40.545&&74.005&112.399&&48.383&74.100\\
			&$PCA$&46.394&71.012&&24.934&39.237&&72.291&109.562&&52.789&80.423\\
			&$RMT$&46.181&70.777&&24.796&39.027&&72.408&109.265&&52.432&80.188\\
			&$LW$&43.444&66.733&&\underline{23.834}&\underline{37.275}&&76.161&114.700&&46.652&71.896\\
			&$LW^F$&63.845&96.725&&62.359&93.102&&78.638&122.279&&60.502&91.636\\
			&$CCC$&43.321&67.023&&27.469&43.276&&75.123&114.858&&44.354&68.515\\
			&$DCC$&\underline{42.585}&\underline{66.060}&&27.081&42.811&&74.082&113.527&&\underline{43.481}&\underline{67.335}\\
			&$DCC^{NL}$&43.739&67.556&&28.016&44.008&&75.278&115.106&&44.766&69.003\\
			
			&$CAB$&\textbf{26.619}&\textbf{41.417}&&\textbf{13.427}&\textbf{20.765}&&\textbf{45.197}&\textbf{69.510}&&\textbf{29.777}&\textbf{46.649}\\
			\\
			\multirow{11}{*}{\rotatebox{90}{$F=60$}}
			&$NA$&44.060&68.125&&26.115&42.593&&72.109&109.679&&47.545&72.605\\
			&$NA^F$&45.268&69.612&&30.016&46.385&&84.049&130.081&&43.879&67.086\\
			&$EWMA$&44.759&68.570&&26.625&41.727&&74.548&112.981&&47.861&73.069\\
			&$PCA$&44.557&68.811&&26.466&43.142&&72.705&111.009&&48.109&73.194\\
			&$RMT$&44.341&68.586&&26.291&42.935&&72.803&110.767&&47.775&72.961\\
			&$LW$&42.043&65.095&&\underline{24.836}&\underline{40.385}&&75.020&113.951&&43.615&66.916\\
			&$LW^F$&62.000&93.988&&62.634&93.568&&69.301&108.405&&59.462&90.061\\
			&$CCC$&42.543&65.895&&28.943&45.693&&68.234&104.807&&43.895&67.683\\
			&$DCC$&\underline{39.404}&\underline{61.909}&&26.764&42.978&&\underline{64.589}&\underline{100.135}&&\underline{40.279}&\underline{63.072}\\
			&$DCC^{NL}$&42.240&65.544&&27.015&43.295&&71.326&108.855&&43.658&67.380\\
			
			&$CAB$&\textbf{24.079}&\textbf{37.328}&&\textbf{13.698}&\textbf{21.649}&&\textbf{43.977}&\textbf{68.513}&&\textbf{25.026}&\textbf{38.428}\\
			\\
			\multirow{5}{*}{\rotatebox{90}{$F=90$}}
			&$NA$&43.982&68.545&&30.179&49.646&&76.435&117.049&&43.496&66.694\\
			&$NA^F$&43.314&66.645&&29.713&46.043&&84.231&130.003&&40.232&\underline{61.574}\\
			&$EWMA$&44.761&68.841&&28.925&\underline{45.742}&&73.405&111.893&&46.704&71.306\\
			&$PCA$&44.474&69.186&&30.520&50.092&&76.878&118.284&&44.100&67.288\\
			&$RMT$&44.249&68.966&&30.343&49.925&&77.116&118.077&&43.709&67.029\\
			\multirow{6}{*}{\rotatebox{90}{$F=90$}}
			&$LW$&42.457&66.219&&\underline{28.694}&47.283&&79.715&121.682&&40.544&62.365\\
			&$LW^F$&60.007&91.075&&61.690&92.268&&\underline{64.073}&\underline{100.740}&&57.731&87.486\\
			&$CCC$&43.276&67.199&&32.551&51.513&&68.021&104.421&&43.036&66.546\\
			&$DCC$&\underline{40.134}&\underline{63.219}&&29.870&48.186&&65.191&100.793&&\underline{39.503}&62.039\\
			&$DCC^{NL}$&42.532&66.357&&29.680&48.045&&70.880&108.147&&42.622&66.080\\
			&$CAB$&\textbf{14.661}&\textbf{23.076}&&\textbf{10.115}&\textbf{16.203}&&\textbf{22.040}&\textbf{34.086}&&\textbf{15.465}&\textbf{24.333}\\
			\\
			\multirow{11}{*}{\rotatebox{90}{$F=120$}}
			&$NA$&48.604&75.417&&36.397&59.018&&84.366&128.460&&46.116&70.620\\
			&$NA^F$&\underline{41.664}&\underline{64.124}&&\underline{31.048}&\underline{48.181}&&85.754&131.620&&\underline{35.717}&\underline{54.821}\\
			&$EWMA$&45.361&69.908&&33.555&53.194&&71.959&109.504&&45.282&69.232\\
			&$PCA$&49.129&76.034&&36.715&59.373&&84.764&129.686&&46.813&71.230\\
			&$RMT$&48.881&75.846&&36.519&59.219&&85.065&129.521&&46.371&71.013\\
			&$LW$&47.111&73.194&&34.955&56.785&&87.320&132.670&&43.296&66.530\\
			&$LW^F$&58.219&88.425&&60.101&90.267&&\underline{64.599}&\underline{100.781}&&55.138&83.629\\
			&$CCC$&44.126&68.610&&36.335&57.587&&71.120&108.182&&41.325&64.245\\
			&$DCC$&43.988&68.569&&34.032&54.862&&73.636&111.512&&41.819&64.965\\
			&$DCC^{NL}$&44.025&68.619&&34.051&54.887&&73.690&111.582&&41.864&65.025\\
			&$CAB$&\textbf{12.118}&\textbf{18.831}&&\textbf{9.496}&\textbf{15.248}&&\textbf{17.864}&\textbf{27.435}&&\textbf{12.147}&\textbf{18.653}\\
			\\
			\multirow{11}{*}{\rotatebox{90}{$F=180$}}
			&$NA$&57.494&88.910&&50.085&79.441&&92.730&140.261&&52.044&80.105\\
			&$NA^F$&\underline{38.294}&\underline{58.953}&&\underline{31.839}&\underline{49.444}&&83.328&126.577&&\underline{29.370}&\underline{45.434}\\
			&$EWMA$&50.062&77.060&&43.438&68.309&&76.943&116.140&&46.536&71.363\\
			&$PCA$&58.045&89.474&&50.496&79.979&&92.932&141.148&&52.787&80.592\\
			&$RMT$&57.763&89.322&&50.352&79.846&&93.237&141.034&&52.245&80.416\\
			&$LW$&56.156&86.964&&49.007&77.759&&94.651&143.001&&49.589&76.623\\
			&$LW^F$&54.314&82.579&&54.255&81.962&&\underline{61.748}&\underline{95.188}&&52.188&79.307\\
			&$CCC$&44.844&70.073&&40.368&64.069&&72.299&109.109&&39.755&62.603\\
			&$DCC$&46.364&72.137&&41.041&65.057&&75.312&113.195&&41.391&64.777\\
			&$DCC^{NL}$&46.406&72.191&&41.077&65.104&&75.366&113.267&&41.432&64.830\\
			
			&$CAB$&\textbf{12.828}&\textbf{19.854}&&\textbf{11.180}&\textbf{17.728}&&\textbf{20.372}&\textbf{30.851}&&\textbf{11.702}&\textbf{18.031}\\
			\\
			\multirow{11}{*}{\rotatebox{90}{$F=250$}}
			&$NA$&76.900&117.511&&107.237&163.540&&79.568&120.252&&56.420&86.818\\
			&$NA^F$&\underline{34.000}&\underline{52.376}&&\underline{33.222}&\underline{51.351}&&65.353&99.107&&25.373&\underline{39.432}\\
			&$EWMA$&53.792&82.799&&53.034&82.561&&72.673&109.796&&48.785&75.090\\
			&$PCA$&77.678&118.355&&108.475&165.210&&79.607&120.884&&57.115&87.187\\
			&$RMT$&77.365&118.210&&108.260&165.073&&79.904&120.760&&56.560&87.032\\
			&$LW$&74.952&114.767&&103.983&158.985&&80.888&122.122&&54.369&83.906\\
			&$LW^F$&49.692&75.555&&47.295&71.544&&\underline{45.710}&\underline{70.616}&&52.408&79.599\\
			&$CCC$&43.349&68.233&&42.431&67.189&&58.278&88.611&&39.597&62.977\\
			&$DCC$&40.167&64.257&&38.751&62.682&&52.667&81.284&&37.446&60.321\\
			&$DCC^{NL}$&40.261&66.455&&49.351&77.363&&80.796&119.188&&\underline{22.552}&44.012\\
			&$CAB$&\textbf{18.246}&\textbf{27.828}&&\textbf{29.049}&\textbf{44.148}&&\textbf{14.976}&\textbf{22.681}&&\textbf{12.182}&\textbf{18.727}\\
			\\
		\end{longtable}
		\endgroup
\end{document}